\documentclass[bibyear]{aa}  

\usepackage{graphicx}
\usepackage{txfonts}
\usepackage{hyperref}
\usepackage[capitalize]{cleveref}
\hypersetup{
 colorlinks,
 linkcolor=blue,
 citecolor=blue,
 urlcolor=blue, 
 }

\usepackage{mathtools}
\usepackage{natbib}
\bibpunct{(}{)}{;}{a}{}{,}

\usepackage{color}
\usepackage{soul}

\definecolor{color1}{RGB}{191, 0, 255}

\renewcommand{\vec}[1]{{\mathbf{#1}}}

\newcommand{\LEt}[1]{\textcolor{cyan}{#1}}

\newcommand{\veps}{{\varepsilon}}

\def\d{{\mathrm d}}

\def\om{{\omega}}
\def\Om{{\Omega}}

\begin{document}

\title{Impact of magnetic field gradients on the development of the magneto-rotational instability}

\subtitle{Applications to binary neutron star mergers and protoplanetary disks}

\author{
  Thomas Celora\inst{1,2}
  \and
  Carlos Palenzuela \inst{3,4}
  \and
  Daniele Vigan\`o\inst{1,2,4}
  \and
  Ricard Aguilera-Miret\inst{4,5}
}

\institute{
  Institut de Ci\`encies de I'Espai (ICE-CSIC), Campus UAB, 08193 Cerdanyola del Vallès, Barcelona, Spain
  \and
  Institut d’Estudis Espacials de Catalunya (IEEC), 08860 Castelldefels, Barcelona, Spain
  \and
  Departament de Fisica,  Universitat  de  les  Illes  Balears,  Palma  de  Mallorca, E-07122,  Spain
  \and
  Institute of Applied Computing \& Community Code (IAC3), Palma  de  Mallorca, E-07122,  Spain
  \and
  University of Hamburg, Hamburger Sternwarte, Gojenbergsweg 112, 21029, Hamburg, Germany
}

\date{Received Month Day, Year; accepted Month Day, Year}

\abstract
{The magneto-rotational instability (MRI) plays a crucial role in accretion disk modelling, driving magnetohydrodynamic turbulence and facilitating enhanced angular momentum transport. Notably, MRI is believed to be pivotal in the development of large-scale poloidal magnetic fields during binary neutron star mergers. However, the few numerical simulations that start from weak seed magnetic fields and capture its growth until saturation show the effects of small-scale turbulence and winding but lack convincing evidence of MRI activity.}
{We investigated how the axisymmetric MRI is impacted by magnetic fields with the realistic, complex topologies of the post-merger phase, where field gradients cannot be neglected. This analysis aims to improve our understanding of the conditions under which MRI develops in more realistic astrophysical scenarios.}
{We performed a linear analysis of axisymmetric MRI under these extended conditions and studied the resulting generalized dispersion relation. After deriving the extended MRI criteria, we first applied them to simple analytical disk models. Finally, we analysed the results obtained from a numerical relativity simulation of a long-lived neutron star merger remnant.}
{We find that radial magnetic field gradients can significantly impact the instability, slowing it and suppressing it entirely if sufficiently large. Specifically, we derived modified expressions for the growth rate and wavelength of the fastest-growing mode. 
Evaluating them in the context of binary neutron star merger remnants, we find that conditions that potentially allow the instability to be active and grow on short enough timescales might be reached only in small portions of the post-merger remnant, and only at late times, $t\gtrsim 100$~ms after the merger.}
{Our results indicate that the role of the axisymmetric MRI in amplifying the poloidal magnetic field in the post-merger environment during the first ${\cal O} (100)$~ms is likely limited.}


\maketitle

\section{Introduction}
The magneto-rotational instability (MRI) is a crucial mechanism for sustaining magneto-hydrodynamic (MHD) turbulence in accretion disks, leading to enhanced angular momentum transport \citep{Shakura+1973,BalbusHawleyRev1998}. 
The instability arises from the interaction between a weak magnetic field and a sheared background flow. 
Its significance in accretion disk modelling was first recognized in the early $1990$s by \citet{BalbusHawley1}, building on earlier work by \citet{Chandrasekar1960} and \citet{velikhov1959}. 
The relevance lies in the fact that, while a differentially rotating disk is hydrodynamically stable according to the Rayleigh stability criterion \citep{Rayleigh1917}, the presence of a magnetic field destabilizes any configuration in which the angular velocity profile decreases radially outwards.

Due to the general nature of the instability and its rapid growth rate, the MRI is often considered a key mechanism in the context of compact binary mergers \citep{shibata2015numerical,radice2024turbulence,ETbluebook}. It is frequently invoked to explain the amplification of magnetic field strengths from the $10^{8}$-$10^{11}$ G typically observed in old neutron star systems \citep{Lorimer2008review} to the `magnetar-level' range of $10^{15}$-$10^{16}$ G, hence aiding in the collimation of relativistic jets and the production of associated gamma-ray bursts \citep{ruiz2016BNSjet,ciolfi2019_100msBNS,Ciolfi2020_BinBNS,CombiSiegel2023,MattiaResRelJet,jiang2025LongTermBImpact,KalinaniJetNoatmo}. 
Other mechanisms are also believed to play a role: magnetic winding, driven by differential rotation, is expected to linearly amplify the toroidal magnetic field, while the Kelvin-Helmholtz instability (KHI)---which develops in the shearing layer at the slip-line between merging neutron stars---has been shown to be highly efficient in amplifying the magnetic field in the core of the remnant \citep{priceRosswog2006,anderson2008magnetized,Kiuchi2015KHI,Kiuchi2018paper}.

Nonetheless, the MRI is expected to play a crucial role in this scenario \citep{duez2006collapse,siegel2013MRI,RoleTurbRicard2023,MinitMRI}, as magnetic winding leads to linear growth; the KHI is believed to be less efficient in the envelope of the remnant.
Recent evidence also suggests that an $\alpha\Omega$-type dynamo operates in the post-merger remnant, contributing to magnetic field amplification at large scales \citep{Kiuchi_dynamo24,most2023dynamoimpact}.
Given its characteristics, the MRI is the natural candidate for sustaining turbulence development in the remnant's envelope and inducing the large-scale dynamo. 
A key feature of the MRI is the presence of a critical wavelength, $\lambda_{MRI} \approx 2\pi v_A / \Omega$, which scales with the Alfvén speed and corresponds to the wavelength at which the instability's growth rate is maximized. As a result, discussions of the MRI in the context of binary mergers often focus on this fastest-growing mode \citep{siegel2013MRI,Kiuchi2018paper,ciolfi2019_100msBNS,MiquelEffMRI}. Demonstrating that this critical wavelength is adequately resolved in simulations ensures that the simulation has sufficient resolution to capture the underlying dynamics.
Notably, although this criterion originates from linear MRI theory, it has proven useful for empirically assessing the non-linear MRI development, as demonstrated by convergence studies in non-linear disk simulations \citep[e.g.][]{HawleyConvergence2013,HawleyConvergence2011}.

However, it is important to remember that the MRI is derived under certain assumptions, which raises questions about the validity of its predictions when those assumptions are violated. For instance, \citet{CeloraMRI} demonstrate that the MRI phenomenology is highly sensitive to the assumption of global axisymmetry. Despite the MRI being based on a plane-wave analysis---suggesting that the instability is local and independent of the large-scale conditions---the assumption of global axisymmetry is hard-wired in the derivation. In fact, \cite{CeloraMRI} show that relaxing this assumption of global symmetry results in a drastic change in the instability's behaviour: the magnetic field acts to suppress or slow the unstable modes in an anisotropic manner.
In this sense, a word of caution is in order. 

Another important piece of the puzzle comes from the long-lived binary neutron star merger simulations presented  in \cite{RoleTurbRicard2023,DelayedJet24}. 
These simulations failed to provide clear evidence of poloidal growth after the vigorous amplification due to the Kelvin-Helmholtz unstable phase, despite the authors demonstrating that they could adequately resolve the supposed wavelength of the fastest-growing mode throughout the simulation.
It is also worth mentioning that such simulations employ large-eddy simulation (LES) schemes \citep{vigano_gradientMHD,carrasco2020gradient,viganoGRMHDLES}
designed to resolve the small-scale dynamics and its impact on magnetic field amplification---see \citet{CeloraHigherLevel,CeloraLagrangFilter,CeloraFibrLES} for recent progress towards the first covariant formulation for relativistic LESs \citep{ETbluebook}.

While this may initially seem surprising, it is relatively easy to come up with a possible explanation. 
The simulations show a magnetic field configuration that remains turbulent even at late times ($t\gtrsim 200$ ms after merger), characterized by a small-scale dominated, randomly oriented field\LEt.
In this scenario, the perturbations will likely undergo a random-walk process, which must be taken into account when calculating the effective Alfv\'en speed \citep{Kiuchi2018paper,TurbBampBNS_2022}. 
As a consequence, the estimate for the Alfv\'en velocity is significantly reduced, which in turn lowers the wavelength of the fastest-growing mode, making the instability even more challenging to resolve numerically. Moreover, in contrast with protoplanetary disks, in such a turbulent highly rotating scenario, the advective timescales can easily be faster than or comparable to the growth timescales, thus challenging the very basic assumption about having a clear background.

The goal of this work was to examine more deeply the insights provided by the above simulations and, for the first time, explore how MRI conditions are influenced by a complex magnetic field.
The standard MRI derivation in fact assumes a weakly magnetized fluid configuration, where the magnetic field is regular enough that we can neglect magnetic field gradients. In this respect, a few works have explored different assumptions: for instance, \citet{dougan2012mri} considered a diamagnetic Ohm’s law but still with relatively simple configurations. 
In this study we focused in particular on the original MRI and the spectrally unstable axisymmetric modes.
Notice also that depending on the net magnetic flux, non-axisymmetric modes may be more relevant to the MRI dynamics and developments \citep[see e.g.][]{BalbusHawley4,Rincon2007_SSPdynamo,Squire2014,Gogichaishvili2017,Held2022}.
We leave the exploration of the impact of complex field geometries on the development of transiently growing non-axisymmetric modes to future studies.
We also present a comprehensive diagnostic of the generalized instability in the post-merger environment, achieved by directly post-processing the numerical output from said simulations. 
This is evidently motivated by our interest in understanding the fineprints of the magnetic field amplification in the post-merger environment, though the results we present are also of broader significance.
As we demonstrate, magnetic gradients can significantly affect the instability, potentially suppressing it entirely if they are large enough. The relevance of this extends beyond just post-merger environments.

The paper is laid out as follows: In Sect. \ref{sec:extended_MRI} we derive the generalized MRI, first discussing the analytical set-up in Sect. \ref{subsec:WKB} and then focusing on the (most relevant) case with radial magnetic field gradients in Sect. \ref{subsec:radialgrads}.
In Sect. \ref{sec:ToyModels} we discuss the generalized instability results for two magnetic field toy models inspired by accretion disk models. 
Finally, we present a detailed diagnostic of the generalized instability in a compact binary merger simulation in Sect. \ref{sec:BNSdiagnostic}, and draw our conclusions in Sect. \ref{sec:conclusions}.

\section{Impact of magnetic gradients on the axisymmetric MRI}\label{sec:extended_MRI}

As a first step, we performed a linear perturbation analysis on a magnetized fluid configuration, considering gradients in both the angular velocity and the magnetic field. 
In particular, we focused on the axisymmetric MRI modes driving the formation of so-called channel modes \citep{Goodman1994} and eventually leading to a magnetized turbulent state as these are destroyed by parasitic instabilities \citep[see][and references within]{MiquelEffMRI}.
First, we verified that the results of the standard MRI are recovered when the magnetic field is constant. Then, we extended the analysis by  assuming non-vanishing magnetic field gradients, which modify the characteristic timescale and wavelength of the instability.

\subsection{Analytical set-up and WKB expansion}\label{subsec:WKB}

We considered a plane-wave analysis of perturbations on top of a non-homogeneous background fluid, which allowed us to account for the impact of differential rotation (and entropy gradients in the most general case) on the time-evolution and propagation of MHD modes. 
Effectively this means that every quantity $A$ is split into background contribution plus fluctuations, and the latter are written in terms of a Wentzel–Kramers–Brillouin (WKB)-like expansion of the form
\begin{equation}
    A = a + \delta a = a + \bar\delta\left(\sum_{q=0}\veps^q a_q\right) e^{i\theta(x)/\veps} \;,
\end{equation}
where $x$ stands for the relevant (spatial) coordinates describing the system.
The book-keeping parameter $\bar\delta$ measures the relative magnitude of the background versus perturbations, while $\veps\approx \lambda/L$ is the ratio between $\lambda$, the typical wavelength of the waves, and $L$, which is the typical length-scale over which the wave amplitude, polarization, and wavelength vary \citep{Anile,ThorneBlandford}. 
Using this expansion for each quantity entering the MHD equations, the terms of order $\mathcal{O}(\bar\delta^0, \veps^0)$ are background terms, while those of order $\mathcal{O}(\bar\delta^1, \veps^0)$ represent linear perturbations. 
Along with this WKB-type ansatz the (mathematical) calculation is based on the assumption that some of the background gradients are `steep' in the sense that $\partial_x a = \mathcal{O}(\bar\delta^0, \veps^0)$, while others are not, in the sense that $\partial_x a = \mathcal{O}(\bar\delta^0, \veps^1)$.
Steep background gradients---relative to a fast varying quantity---will consequently affect the perturbation equations (as well as changing the notion of background). 
By expanding the phase $\theta(x)/\veps = \theta(0)/\veps + k_x x + \dots$, we can then work out a dispersion relation where such gradients will appear. 
See \cite{CeloraMRI} for more details of this multi-scale WKB expansion in the context of the MRI.

We next observed that background gradients enter explicitly in the perturbation equations only when they appear as part of a non-linear term in the MHD equations (e.g. $A\partial_x B$). 
Since the MHD equations are linear in the time-derivatives, temporal gradients do not explicitly modify the perturbation equations or the dispersion relation.
In contrast, spatial gradients---both of the magnetic field (which is neglected in the standard derivation) and of the background velocity profile (which is considered in the standard derivation)---do play a direct role. This does not imply, however, that temporal gradients are irrelevant.
The assumption that the background is stationary, or at most evolving slowly, is crucial; without this assumption, the separation of background and perturbation would not make sense. Assuming that the background evolves on a slow timescale $T$, any unstable mode must grow on a much shorter timescale $\tau \ll T$ (or at least $\tau \lesssim T$) to be relevant. If this condition is not satisfied, the instability would not have sufficient time to grow before the background changes.

\subsection{Linear analysis of the MHD equations}\label{subsec:radialgrads}

We next derived the generalized instability in the presence of a non-homogeneous magnetic field. 
We focused on the impact of adding radial magnetic field gradients since, as we will see shortly, they can have a crucial impact on the instability results. 
Our starting point was the non-relativistic ideal MHD equations
\begin{subequations}
\begin{align}
    &\partial_t \rho + \nabla \cdot(\rho \vec v) =0 \;, \\
       & \partial_t \vec v + \vec v\cdot\nabla\vec v = - \frac{1}{\rho}\nabla P - \nabla\Phi + \frac{1}{4\pi \rho}\left[\vec B \cdot\nabla \vec B - \nabla\left(\frac{B^2}{2}\right) \right]  \;,\\
    &\partial_t \vec B = \nabla \times (\vec v \times\vec B)\;,
\end{align}
\end{subequations}
where $\rho,\,\vec v,\,P,\,\Phi$, and$\,\vec B$ indicate the mass density, fluid velocity, thermodynamic pressure, gravitational potential and magnetic field, respectively.
We then performed the analysis using the local co-rotating frame construction presented in \cite{CeloraMRI}, who explicitly show that the axisymmetric MRI can be reformulated this way. 
In short, this means setting up (and working with) a local Cartesian frame (comoving with the background flow) such that the $x$-axis always points in the radial direction, the $y$-axis in the azimuthal one, and the $z$-axis is aligned with the global axis of rotation.
Clearly, such a construction is not necessarily possible in absence of a global sense of rotation (and hence a notion of axis of symmetry).
In this regard, while this formulation is equivalent to the original, it has the advantage of explicitly highlighting the key role that axisymmetry plays in the result.

We started by assuming a circular velocity profile, $\vec v = \Omega(R)R \hat\varphi$, and then considered an observer co-orbiting with the fluid on some given orbit, $R_0$. 
The observer velocity is then $\vec v_{obs}= \Omega(R_0)R\hat \varphi = \Omega_0 R\hat \varphi$ while the background velocity as measured in the co-rotating frame is\footnote{To avoid confusion, let us explicitly state that here $x$ is the coordinate along the x-axis in the local co-rotating frame construction.}
\begin{multline}\label{eq:corotatingObserver}
    \vec v' = (\Omega - \Omega_0) R \hat \varphi  = \frac{\partial\Omega}{\partial R}\Big|_{R_0} (R_0+x)x \hat\varphi \\
    \Longrightarrow \vec v'  = S_0 x \hat y + \mathcal{O}(x^2) \;,\quad S_0 = \frac{\partial\Omega}{\partial \log R}\Big|_{R_0}\;. 
\end{multline}
As a result, when writing the linearized MHD equations in the co-rotating frame, the relevant background velocity gradients to consider are $\partial_i v_j = S_0 \delta_{ix}\delta_{jy}$. 
When it comes to the magnetic field, the one measured by the co-rotating observer is the same as in the inertial frame, as one can easily see using Lorentz transformations and taking the Newtonian limit. 
In essence, we assumed the background field to satisfy
\begin{equation}
    \vec B = \vec B(x) = B^y(x)\hat y + B^z(x) \hat z \;, 
\end{equation}
namely we imposed only radial gradients in this section and assumed no radial field---noting that this is parallelled in the original MRI calculation to disentangle the instability from winding, and is later shown not to change the instability criteria.

We performed a plane-wave analysis with perturbations proportional to $\exp(i k_j x^j - i \omega t)$ under the following simplifications: consider only axisymmetric modes ($k_y=0$), assume that the flow is incompressible ($\delta\rho =0$) and consider wavelengths small compared to the scale height, allowing us to neglect pressure gradients entirely\footnote{Mathematically, this is equivalent to considering a barotropic equation of state, and is advantageous also because the conventional MRI is discussed within the context of the so-called Boussinesq approximation \citep{barletta2022boussinesq}, whose validity in the post-merger environment can be questioned.}.
These simplifications are possible because the key ingredients to reproduce the MRI are a non-zero magnetic field and a sheared background. Finally, we considered the gravitational potential as being externally sourced, so that\footnote{This is also done in the original MRI derivation in the context of accretion disks modelling, where the fluid is not self-gravitating. This simplification is still reasonable in binary neutron star mergers, as the MRI is expected to play a role mainly in the envelope of the merger remnant.} $\delta\Phi=0$.
After imposing all these conditions, the linearized equations for the perturbations can be written as 
\begin{subequations}\label{subeqs:perturbedMHDBgrad}
\begin{align}
    & k_x \delta v^x + k_z \delta v^z = 0 \;, \\
    \label{eq:xEulerMHDBgradR}
    & \begin{multlined}[b]
    -i \om \delta v^x - 2\Om_0\delta v^y + \textcolor{blue}{(\partial_x v_A^y)\delta v_A^y + (\partial_x v_A^z)\delta v_A^z }  \\
    + i k_x (v_A^y \delta v_A^y + v_A^z \delta v_A^z ) -i \delta v_A^x k_z v_A^z = 0 \;.
    \end{multlined} \\
    -i &\om \delta v^y +\frac{\kappa^2 }{2\Om}\delta v^x \textcolor{blue}{- (\partial_x v_A^y )\delta v_A^x} - i k_z v_A^z \delta v_A^y  = 0 \;,\\
    -i &\om \delta v^z  \textcolor{blue}{- (\partial_x v_A^z )\delta v_A^x} + i k_z v_A^y \delta v_A^y = 0 \;, \\
    -i & \om \delta v_A^x - i v_A^z k_z \delta v^x = 0 \;,  \\
    -i &\om \delta v_A^y \textcolor{blue}{+ (\partial_x v_A^y) \delta v^x} - i v_A^z k_z \delta v^y - s_0 \delta v_A^x = 0 \;,\\
    -i & \om \delta v_A^z \textcolor{blue}{+ (\partial_x v_A^z )\delta v^x} -i v_A^z k_z \delta v^z = 0\;,
\end{align}
\end{subequations}
where we introduced the Alfv\'en velocity and the square of the epicyclic frequency: 
\begin{equation}
    v_A^i =\frac{B^i}{\sqrt{4\pi \rho}}\;, \quad \kappa^2 = 2\Omega \frac{\d\Omega}{\d\log R} + 4\Omega^2\;.
\end{equation}
We have also highlighted in blue the additional terms originating from gradients in the background magnetic field. 

For the next steps we followed the strategy adopted in the original derivation. First, we rewrote all perturbations in terms of $\delta v^z$, and introduced 
\begin{equation}
    \partial_x v_A^z= a \partial_x v_A^y\;,
\end{equation}
where $a$ can be positive or negative, to slim down the notation.
Doing this, and solving a two-equation linear system for $\delta v_A^y $ and $\delta v^y$ we arrive at  
\begin{subequations}\label{subeqs:substitutionsBgrad}
\begin{align}
    & \delta v^x = - \frac{k_z}{k_x}\delta v^z \;,\\
    & \delta v^y = i \left(\frac{\om}{\omega^2 - p^2}\right)\frac{k_z}{k_x}\left(\frac{\kappa^2}{2\Om} - S_0\frac{p^2}{\om^2} \right)\delta v^z \;,\\
    & \delta v_A^x = \frac{k_z}{k_x} \frac{p}{\omega}\delta v^z \;,\\
    & \begin{multlined}[b]
        \delta v_A^y = -i \frac{k_z}{k_x}\frac{1}{\om^2 - p^2 }[2p\Om  \textcolor{blue}{ - \frac{\partial_xv_A^y}{\om}(\om^2 - p^2)}] \delta v^z \\
        = \frac{1}{v_A^y}\left(\frac{\om}{k_z}\textcolor{blue}{-ia\frac{\partial_x v_A^y}{k_x}\frac{p}{\om}}\right)\delta v^z \;,
    \end{multlined} \\
    & \delta v_A^z = - \left(\frac{p}{\om}\textcolor{blue}{-ia\frac{\partial_x v_A^y}{\om}\frac{k_z}{k_x} }  \right)\delta v^z \;,
\end{align}
\end{subequations}
where we have introduced the shortcut $p \equiv k_z v_A^z$.
Substituting these into the x-component of the Euler equation above (cf. \cref{eq:xEulerMHDBgradR})  and rearranging terms, we arrive at the following dispersion relation:
\begin{equation}\label{eq:disprelBgrad}
    a_0 \om^4 + a_2 \om^2 + a_4 = 0\;,
\end{equation}
where
\begin{align*}
   a_0 &= 1 \textcolor{blue}{-i \frac{\partial_x v_A^y}{v_A^y}\frac{k_x}{k_z^2}} \;, \\
   a_2 &=\begin{multlined}[t]
       - \Big[2p^2 + \kappa^2 \frac{k_z^2}{k^2} \textcolor{blue}{-i a(\partial_x v_A^y) \frac{k_xk_z}{k^2}p -i \frac{(\partial_x v_A^y) }{v_A^y}\frac{k_x}{k^2}p^2}
    \\
    \textcolor{blue}{+a\frac{(\partial_x v_A^y) ^2}{v_A^y}\frac{k_z}{k^2}p - a^2(\partial_x v_A^y) ^2 \frac{k_z^2}{k^2}}\Big] \;,
   \end{multlined}  \\ 
    a_4 &= \begin{multlined}[t]
        \Big[p^4 + 2\Om s_0 \frac{k_z^2}{k^2}q^2 \textcolor{blue}{-i a (\partial_x v_A^y)  \frac{k_xk_z}{k^2}p^3 +a\frac{(\partial_x v_A^y) ^2}{v_A^y}\frac{k_z}{k^2}p^3}\\
        \textcolor{blue}{- a^2 (\partial_x v_A^y) ^2 \frac{k^2_z}{k^2}p^2}\Big] \;.
    \end{multlined}
\end{align*}
This dispersion relation is a polynomial of fourth order as in the standard MRI analysis, but with much more involved coefficients that depend not only on the angular velocity gradients but also on the magnetic field ones.

\subsection{Recovering the standard MRI: Constant magnetic fields}\label{subsubsec:MRI}
As a first step, let us show that the usual MRI can be recovered from the more general dispersion relation \cref{eq:disprelBgrad}. 
This provides a sanity check and also serves as a comparison for the more general results that we will discuss shortly.
To recover the standard MRI we disregard the terms originating from the gradients of the magnetic field in \cref{eq:disprelBgrad} (i.e. the blue terms).
We also considered purely vertical modes (that is set $k_x=0$), which are the ones that grow the fastest.
With these simplifications, we arrived at the following dispersion relation: 
\begin{equation}
    \om^4 - (2p^2 + \kappa^2)\om^2 + p^2 (p^2 + 2\Om S_0) = 0\;, 
\end{equation}
which is a real, quartic algebraic equation for $\om$.  From this equation it is straightforward to check that, in the limit $p^2\gg \kappa^2 \sim \Om s_0$, the solutions are $\om^2 = p^2$ and there is no instability. 

The solution for $\om^2$ of the full dispersion relation is given by
\begin{equation}
    \om^2_{\pm} = \frac{1}{2}\left[2p^2 + \kappa^2 \pm \sqrt{\kappa^4 + 16 \Om^2 p^2}\right]\;.
\end{equation}
Notice that the $\om^2_\pm$-solutions are real since the discriminant is always positive. 
However, if one of these solutions is negative, the corresponding two $\om$-solutions will be complex conjugates, and hence one is unstable. 

To derive the instability criterion, one can then study the marginal case $\om^2 \approx 0$, in which one can neglect the $\om^4$ term and solve the dispersion relation as
\begin{equation}
    \frac{\om^2}{p^2} = \frac{p^2 + 2\Om S_0}{2p^2 + \kappa^2} \;.
\end{equation}
We could restrict the analysis to `small wavenumbers', $p^2\ll \kappa^2\sim\Om s_0$, as we have already found that in the opposite limiting case there is no instability.
The denominator is always positive, since the Rayleigh criterion (ensuring hydrodynamic stability) is precisely $\kappa^2 >0$ \citep{Rayleigh1917}. 
It is then easy to check that the sign of $\om^2$ is dictated by the sign of $S_0$.
In essence, we have recovered the standard MRI criterion, which states that, if the background angular velocity decreases radially outwards (i.e. $S_0<0)$, then the system is unstable. 
It is worth emphasizing that unstable modes can still exist when $p\sim\kappa$, although in this case the precise instability criterion is no longer wavenumber-independent.

The fastest vertical growing mode is found by considering that the $\om^2_-$-solution is a function of $k$ and solving for the stationary points, namely
\begin{equation}
    \frac{d\om^2_-}{dk} = 0 \Longrightarrow 
    k_{MRI} = \left(\frac{B^z}{\sqrt{4\pi \rho}}\right)^{-1}\left(\Om^2 - \frac{\kappa^4}{16 \Om^2}\right)^{1/2} = \frac{2\pi}{\lambda_{MRI}} ,
\end{equation}
which allow us to define $\lambda_{MRI}$.
We could also work out the corresponding growth rate:
\begin{equation}\label{eq:stdMRItau}
    \om^2_{MRI} = \om^2_-(k_{MRI}) = -\frac{1}{4}S_0^2 \Longrightarrow \tau_{MRI} = \frac{2}{|S_0|} \;,
\end{equation}
which sets the timescale of the instability $\tau_{MRI}$ as the e-folding time of the fastest-growing mode.   
Although all these results are obviously well known in the literature \citep{BalbusHawleyRev1998}, it is worth stressing a point that is relevant in particular for numerical simulations: since the wavenumber of the fastest-growing mode scales with the inverse of the background magnetic field, the instability occurs at larger wavelengths and is easier to resolve when the background magnetic field is stronger (i.e. $\lambda_{MRI} \propto B^z$). 
In contrast, the precise value of the magnetic field does not influence the instability criterion nor the growth rate of the fastest-growing mode. 

Before we move on to the case with radial gradients, it is important to recall that a fundamental assumption of the linear analysis is that unstable wavelengths should be much smaller than the length-scales over which pressure, density and gravity fields vary. 
This is relevant because it implies that wavenumbers smaller than the minimum value $k_{min} = 2\pi/L_{disk}$, where, say, $L_{disk}={\rm min}(R,H)$ and $H$ is the scale-height while $R$ is the disk size, cannot be excited in practice. 
The instability then operates only when the wavenumber exceeds this threshold $k>k_{min}$, which represents a consistency condition ensuring that the mode fits within the system. On the other hand, it is suppressed in regimes where $p\gg\kappa$, implying that, for a given background magnetic field, there exist a maximum wavenumber $k_{max}$ for which the MRI is active.  
The net effect, a well-established result, is that sufficiently strong magnetic fields suppress the instability by preventing both conditions from being simultaneously satisfied.
Similarly, the growth timescale should be much smaller than the timescale of variation of the background field. While this likely poses no problems in scenarios like protoplanetary disks, we have to be very careful in the case of the fast-evolving dynamics of a binary merger remnant.

\subsection{Extended MRI with radial gradients of the magnetic field}\label{subsubsec:MRIBgrads}
We solved the dispersion relation by retaining the terms that originate from background gradients in the magnetic field. In this case, the dispersion relation is still a quartic algebraic equation but no longer real since the coefficients are generically complex-valued. 
Nevertheless, we could consider purely vertical modes as in the previous section: by setting $k_x=0,$ the dispersion relation reduces to a real quartic algebraic equation again,%
\begin{equation}\label{eq:BofRdisprel}
    \om^4 - \left[2p^2 + \kappa^2 + f \right] \om^2 +p^2 \left[p^2 + 2\Om S_0 + f \right] = 0  
,\end{equation}
where all the information about the gradients of the magnetic field is contained in the new quantity defined as
\begin{equation}\label{eq:def_f}
    f \equiv  (\partial_xv_A^y)(\partial_xv_A^z)\frac{v_A^z}{v_A^y} - (\partial_xv_A^z)^2 \;. 
\end{equation}
To understand the impact of the new terms, we followed the same strategy adopted in the standard analysis. 
Solving \cref{eq:BofRdisprel} for $\om^2$ we arrive at 
\begin{equation}\label{eq:Omsqsol_gradients}
    \om^2_\pm =\frac{1}{2}\left[ 2p^2 + \kappa^2 + f \pm \sqrt{(\kappa^2 + f)^2 + 16 \Om^2 p^2 }\right] \;
\end{equation}
and observe that, as before, the discriminant is positive: the $\om^2$-solutions are real, implying that there is an instability if one of them is negative. 
Following the same procedure as in the previous subsection, we focused on the $\omega_-^2$ solutions as this family contains the (potentially) unstable modes with larger growth rates. We could then compute the wavenumber of the fastest-growing mode and the corresponding maximum growth rate, namely
\begin{align}
     k_{eMRI} &=  \left(\frac{B^z}{\sqrt{4\pi \rho}}\right)^{-1}\left(\Om^2 - \frac{(\kappa^2 + f)^2}{16 \Om^2}\right)^{1/2} = \frac{2\pi}{\lambda_{eMRI}}\;, \label{eq:RadialGradBFastest_omega}\\
     \om^2_{eMRI} &= -\frac{1}{4}\left(s_0 - \frac{f}{2\Om}\right)^2  \Longrightarrow
     \tau_{eMRI} = \frac{2}{|s_0 - f/2\Om|} \;.
     \label{eq:RadialGradBFastest_tau}
\end{align}
Regarding the instability criterion, we first observe that the regime where $p^2\gg f,\kappa^2\sim\Om s_0$ leads again to the same equation as in the standard MRI case, namely
\begin{equation}
    \om^4 - 2p^2 \om^2 + p^4 = (\om^2 - p^2)^2 = 0 \Longrightarrow \om = \pm (\vec v_A \cdot\vec k)\;.
\end{equation}
Therefore, in this regime there is no instability and the modes are essentially Alfv\'en waves\footnote{This is consistent with the fact that we have effectively considered a sound-proof limit, where fast magneto-sonic waves are filtered out while slow magneto-sonic waves reduce to Alfv\'enic ones \citep{Vasil_2013}.}. 

To derive the modified instability criterion, we again considered the marginal case, $\om^2 \approx 0$, and restricted to the regime\footnote{Also in this case we may have unstable modes also in the regime where $p^2$ is commensurate to $\Om S_0$ or $f$, but once again the instability criterion is no longer independent of $p$.} $p^2 \ll f, \kappa^2$:
\begin{equation}\label{eq:RadialGradBMarginal}
    \frac{\om^2}{p^2} = \frac{p^2 + 2\Om S_0 + f}{2p^2+ \kappa^2 + f} \simeq \frac{ 2\Om S_0 + f}{ \kappa^2 + f}  \;.
\end{equation}
To better understand the impact of the additional gradients on the instability, we considered different choices that lead to increasing complexity in the analysis. 
In the first case we considered radial gradients only in the azimuthal magnetic field (i.e. $\partial_x v_A^z=0$, $\partial_x v_A^y\neq 0$). With this setting, the function $f$ vanishes trivially, implying that the phenomenology is the same as in the standard MRI.

In the second case we included radial gradients only in the vertical magnetic field (i.e. $\partial_x v_A^z\neq 0$, $\partial_x v_A^y= 0$). In this case, $f=- (\partial_x v_A^z)^2 <0,$ and from \cref{eq:RadialGradBFastest_tau} it is clear that the timescale of the instability is longer than in the standard MRI case. 
As for the instability criterion, from \cref{eq:RadialGradBMarginal} we immediately obtain\footnote{We have here re-introduced global cylindrical coordinates, in order to make the instability condition more transparent.}
\begin{multline}
    |f| < \kappa^2 \longrightarrow \left(\frac{\partial v_A^z}{\partial R}\right)^2 < 2\Om \frac{\d\Om}{\d\log R} + 4\Om^2\\
    \Longrightarrow \left(\frac{\partial v_A^z}{\partial R}\right)^2 < \Om^2 \;\; \text{for a Keplerian disk}\;.
\end{multline}
In this set-up the instability window gets smaller, as magnetic field gradients slow down the instability and can even stabilize the system if they are large enough. Consequently, there is an increase in $\tau_{eMRI}$, which means that the instability grows more slowly, compared to the standard case. 

Finally, we considered the general case, where both magnetic field components have non-vanishing radial gradients  (i.e. $\partial_x v_A^z\neq 0$, $\partial_x v_A^y \neq 0$). This case might be better understood by introducing the (signed) characteristic scales over which the background magnetic field varies:
\begin{equation}\label{eq:ChGradientScalesDef}
    \frac{\partial_x v_A^y}{v_A^y} = \frac{1}{L_y} \;, \quad \frac{\partial_x v_A^z}{v_A^z} = \frac{1}{L_z} \Longrightarrow f= -(\partial_x v_A^z)^2 \left(1 - \frac{L_z}{L_y}\right)\;.
\end{equation}
This is convenient as it allows us to distinguish three scenarios. 
A first scenario is when $L_y \gg L_z$, that is, when gradients in the azimuthal component are much smaller than those in the vertical field. 
In this situation we effectively get back to the phenomenology just discussed for the case where $f = -(\partial_x v_A^z)^2$. 
A second scenario is when $L_z \gg L_y$, namely vertical gradients are negligible when compared to azimuthal ones. 
In this case, we can rewrite $f= (v_A^z)^2/ L_yL_z $, and observe that its sign is dictated only by the relative orientation of the radial gradients in the two components. 
The third, most general scenario is the one in which the two lengths are comparable $L_y \approx L_z$, and the sign of $f$ is dictated by the relative orientations as well as the magnitudes of the gradients of the two magnetic field components. 

After presenting all the possible scenarios, the generalized instability can be characterized as follows depending on the sign of the f-function:
\begin{enumerate}
    \item If $f>0$: the instability window is defined by $f < 2\Om |S_0|$ (for a Keplerian disk $f< 3\Om^2$), while $\lambda_{eMRI}$ grows together with a decrease in  $\tau_{eMRI}$.\label{eMRIcriterion_fpos}
    \item If $f<0$: the instability window is given by $|f|<\kappa^2$ (for a Keplerian disk $|f|< \Om^2$), while $\tau_{eMRI}$ grows. The wavelength $\lambda_{eMRI}$ can either decrease or increase.\label{eMRIcriterion_fneg} 
\end{enumerate}
Although $\lambda_{eMRI}$ and $\tau_{eMRI}$ can shift towards larger or smaller values depending on the sign of $f$, we find that radial gradients generically reduce the instability window and can make the system stable for sufficiently large values, of the order of $\mathcal{O}(\Om^2)$ for a Keplerian disk.

Notice that we directly considered the case where $S_0 <0$, as this is the regime where the instability appears in the standard MRI setting. 
It is interesting to notice that the case $S_0>0$  can be unstable in the presence of radial magnetic field gradients, provided that
\begin{equation}\label{eq:s0positivecase}
    f<0 ~~~, ~~~~ \kappa^2 - 4\Om^2 \leq|f|\leq \kappa^2 \;.
\end{equation}
Note that, in the practical cases presented in the following sections, we have not found any region satisfying these quite restrictive conditions.

Finally, while it is natural to consider the regime $p^2 \ll f,\,\kappa^2$ to derive a simple instability criterion and to compare with the standard scenario, it is worth revisiting the issue of consistency with the basic assumption of linear analysis---namely, that unstable wavelengths should be smaller than the relevant background variation scales. We can distinguish two qualitatively different regimes. In absence of significant gradients, compatibility with the basic assumption of linear analysis together with the fact that the instability is suppressed when \( p^2 \gg \kappa^2 \), implies that very strong magnetic fields can inhibit the instability (as discussed previously).
In the presence of strong background gradients, the instability is suppressed when $p^2 \gg \kappa^2, f$. In this case, if $f \gtrsim \kappa^2$, the relevant maximum wavenumber becomes $k_{\max} \simeq 1/L_B$, where $L_B$ is the characteristic length-scale of magnetic field variations. Here, consistency with the basic assumption of linear analysis does not lead to any additional condition depending on magnetic field strength.
Ultimately, this simply reflects the fact that, in the presence of gradients, an additional characteristic frequency, \( \sqrt{f} \), enters the problem, leading to a richer and more varied instability phenomenology.

\subsection{MRI with generic gradients}\label{subsec:recap}
We have just demonstrated that radial magnetic field gradients can significantly impact the MRI criterion and its phenomenology. This raises the natural question of whether we can extend the analysis to study the effect of, for example, vertical magnetic field gradients.
In principle, we could aim to consider the most general setting, though this would inevitably complicate the resulting dispersion relation, making it less intuitive and harder to interpret. Instead, the strategy we adopted is to examine a series of different settings and analyse each of them individually.
We omit a detailed analysis here (which can be found in \cref{app:verticalgrads}) and instead provide a summary of the results in \cref{tab:recap}. This is useful as it clearly highlights the central role that radial magnetic field gradients play in generalizing the standard MRI criteria to complex field configurations.
As a result, we focus in the following sections on exploring the consequences of the generalized criterion we have just derived. 
\begin{table}[]
    \centering
    \caption{Impact of adding the vertical and radial gradients in the background quantities to the MRI phenomenology.}
    \label{tab:recap}
    \begin{tabular}{c|c|c|c|c}
        \hline
        $\partial_R \Om$ & $\partial_z \Om $ & $\partial_R \vec B$ & $\partial_z \vec B$ & impact on MRI \\
        \hline 
        \hline
        yes & no & no & no & standard/reference case \\
        yes & yes & no & no & none \\
        yes & no & no & yes & none \\
        yes & yes & no & yes & none \\
        yes & no & yes & no & impact on $\lambda,\tau$ + reduced window\\
        yes & yes & yes & no & limits to case with only $\partial_R\vec B$ \\
         \hline
    \end{tabular}
\end{table}

\subsection{MRI-driven dynamos and axisymmetric MRI}
Throughout this work, and in the preceding discussion of this section, we have focused on the influence of magnetic field gradients on axisymmetric, spectrally unstable MRI modes. 
These modes are particularly relevant in scenarios where, say, the accretion disk is threaded by a weak mean vertical magnetic field, such that there is a net poloidal magnetic flux. 
In such cases, axisymmetric MRI modes become unstable, giving rise to so-called channel modes, which are eventually disrupted by secondary parasitic instabilities leading to the magnetized turbulent state \citep{Goodman1994,MiquelEffMRI}.
It is important to note, however, that MRI-driven dynamos have also been extensively investigated in alternative set-ups, such as the toroidal magnetic flux and the zero-net-flux configuration \citep[see e.g.][]{FromangZNF2007,KapylaZNF2010,Herault2011,Mamatsashvili2013,Gogichaishvili2017,Held2022}. These studies are designed to explore the (long-term) sustainability of a turbulent magnetized state in the absence of a mean poloidal magnetic field, as might be the case in certain accretion disk environments.
In particular, it is known that a subcritical, self-sustaining dynamo process is possible in such configurations. This process fundamentally relies on the non-modal (transient) amplification of non-axisymmetric modes \citep{Rincon2007_SSPdynamo,Mamatsashvili2013,rincon2019dynamo}.
Our present study does not address the influence of additional gradients on the development of such non-axisymmetric modes, which would require a more global approach than the local WKB analysis adopted here.

\section{Application to analytical models}\label{sec:ToyModels}
In this section we explore the phenomenology of the extended MRI criteria using simple analytical models for the magnetic field configuration, which are inspired from the planetary and accretion disks literature (e.g.  \citealt{pelletier1992hydromagnetic}). 

\subsection{Vertical magnetic field with radial power-law decay}\label{subsec:AnalyticalVertical}
We considered a simple magnetic field configuration with a purely vertical magnetic field with a constant and a decaying components, namely
\begin{equation}\label{eq:analyticBverticalpowerlaw}
    B^z =  B_0 + \frac{k}{R^n} = B_0 + \frac{B_1}{x^n} \;, \quad x = \frac{R}{R^*} \;,
\end{equation}
where $n$ sets the power-law decay and $R^*$ is some reference radial distance. 
The redefinition in terms of $x$ ensures that $B_1$ has the same dimensions as $B_0$, and also that the magnetic field at $R^*$ is the same across all models with different values of $n$.
In this configuration the characteristic scale of the magnetic field gradients reads
\begin{equation}
    L_z = - \frac{xR^*}{n}\left(\frac{B_0}{B_1} x^n +1\right)\approx -2\frac{R^*}{n}\;,
\end{equation}
where, in the second step, we have assumed: (i) that the magnitudes of the decaying and constant components are comparable, and (ii) that $x \approx 1$. 
This expression shows that $L_z$ is primarily set by $R^*$ and scales inversely with the power-law exponent $n$---larger values of $n$ correspond to steeper magnetic field gradients. In the limit where $B_0 \gg B_1$, the radially varying component becomes negligible compared to the constant offset, and the characteristic scale becomes much larger, $|L_z| \gg 1$, accordingly.
We assumed that the velocity field follows a Keplerian profile, $\Omega = c_\Omega R^{-3/2}$, and introduced the dimensionless parameter
\begin{equation}
    C = \frac{B_1^2 R^*}{4\pi\rho c_\Omega^2}\;, 
\end{equation}
which quantifies the relative strength of the radially varying magnetic component with respect to the shear in the background flow. We considered values $C\lesssim 0.1$, corresponding approximately to a situation in which the total magnetic energy is less than 10\% of the kinetic energy, for which the magnetic feedback on the hydrodynamics can be safely neglected.

Considering the definition of $f$, \cref{eq:def_f}, it is evident that $f<0$, in which case the generalized instability criterion reads
\begin{equation}
    \frac{-f-\kappa^2}{\Om^2} = n^2 \,C\, x^{1-2n} -1 < 0\;.
\end{equation}
It is straightforward to compute the timescale of the instability $\tau_{eMRI}$, \cref{eq:RadialGradBFastest_tau}, and compare it to the standard MRI result $\tau_{MRI}$, \cref{eq:stdMRItau}:
\begin{equation}
    \frac{\tau_{MRI}}{\tau_{\text{eMRI}}} =  \frac{|2\Om s_0 -f|}{|2\Om s_0|} = \left|1 - \frac{1}{3}n^2 \,C\, x^{1-2n}\right| \;.
\end{equation}
We also compared the wavelength of the fastest-growing mode, obtaining
\begin{equation}\label{eq:verticalAnalyticallambda}
    \frac{\lambda_{MRI}}{\lambda_{eMRI}} = \sqrt{\left|1 -\frac{1}{15}\left[1 - (1- n^2\,C\,x^{1-2n})^2\right]\right|}\;.
\end{equation}
It is not surprising that $B_0$ does not enter into the formula for the generalized instability criterion nor that of the timescale of the instability, as these depend only on the gradients of the magnetic field.
This is analogous to the standard MRI case, where the instability criterion makes no reference to the background magnetic field \citep{BalbusHawleyRev1998}. 
We also stress that we here focused on the wavelength and growth rate of the fastest-growing mode, as this provides a straightforward basis for comparison with the standard case. It is important to note, however, that this mode is most relevant only when its wavelength remains sufficiently smaller than the characteristic size of the system. If this condition is not satisfied, slower-growing modes with more suitable wavelengths may dominate the dynamics. Nevertheless, the comparison remains meaningful, since the function $f$, which captures the effect of radial gradients, enters the dispersion relation in a `wavenumber-independent' manner (cf. \cref{eq:Omsqsol_gradients}). As a result, its impact can be extrapolated to modes beyond the fastest-growing one.

\begin{figure*}
    \centering
    \includegraphics[width=17cm]{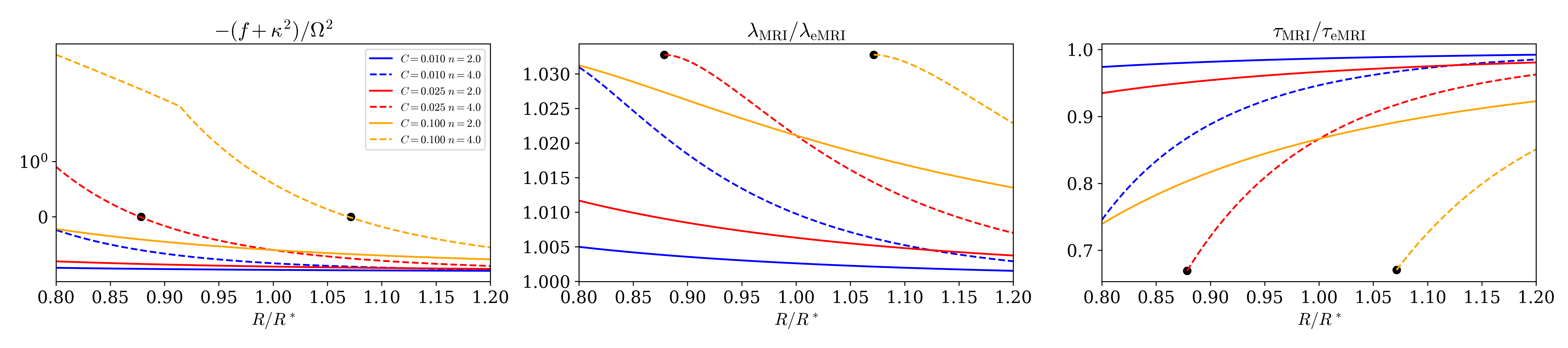}
    \caption{\textit{Extended MRI analysis for the analytical model given by a vertical magnetic field with radial power-law decay in \cref{eq:analyticBverticalpowerlaw}}. \textit{Left}: Generalized instability criterion. \textit{Middle}: Comparison between the generalized and standard MRI for the wavelength of the fastest-growing mode. \textit{Right}: Timescale of the instability. We have considered the values with $n=2,\,4$ and $C=0.01, \,0.025,\,0.1$ in the model defined by \cref{eq:analyticBverticalpowerlaw}.  The wavelength and timescale are shown only when the instability is active according to the generalized criterion, a transition that is marked with black dots.}
    \label{fig:verticalAnalytical1D}
\end{figure*}

It is illustrative to plot the quantities that we just derived considering different values of $n$ and $C$, starting with the generalized instability criterion, displayed in the left panel of \cref{fig:verticalAnalytical1D}. 
As shown in the figure, the impact on all three plotted quantities increases with larger values of the constant $C$ and steeper power-law decays---parameters that together determine the strength of the gradients. In particular, for $n = 2$, the gradients are never strong enough to suppress the instability (given the values of $C$ considered), although they do lead to changes of a few percent in the characteristic wavelength and up to $20\%$ in the growth timescale. This changes for higher values of $n$ (and larger values of $C$), where the gradients become strong enough to suppress the instability in certain regions of the domain. Most notably, we observe a significantly stronger effect on the growth timescale, with reductions reaching up to $30\%$ in the most extreme cases.

\subsection{Vertical and toroidal magnetic field}\label{subsec:Analytical1OverR}
A second case that is worth considering is motivated by the simplified analytical models of disks \citep{galli06,jafari19}. Protoplanetary disks are an excellent observational probe of the magnetic field topology: polarization measurements of high-resolution millimetre observations (e.g. \citealt{beltran19,maury22,huang24}) provide the magnetic field direction on the plane of the sky, which can be often approximated as a `hourglass' configuration, i.e. a vertical field pinched on the mid-plane. Neglecting the turbulent contributions, which are actually likely relevant in several cases \citep{huang25}, the field can be approximated by non-vanishing vertical and azimuthal components. 

To investigate the effects of non-vanishing radial gradients in both magnetic field components, we introduced two key quantities: the ratio ($r$) between the characteristic gradient scales in the vertical and azimuthal components, and a dimensionless parameter, $\tilde C,$ which characterizes the strength of the radial gradient in the vertical Alfvén speed relative to the background shear, namely
\begin{equation}
    r = \frac{L_z}{L_y} ~~,~~
    \tilde C = \left(\partial_R v_A^z\right)^2 \frac{{R^*}^3}{c_\Omega^2}\; \Longrightarrow{ f =  \tilde C (r-1)}\;,
\end{equation}
We also emphasize that, since the characteristic scales carry a sign (cf. \cref{eq:ChGradientScalesDef}), their ratio can be positive when the two magnetic field components vary in the same radial sense (either both increasing or both decreasing), and negative when they vary in opposite senses.
In the case of a vertical magnetic field obeying a radial power-law profile, as considered in the previous section, this quantity is related to the parameter $C$ via $\tilde C \approx n^2 C$. We considered the range $\tilde C \in [0, 0.24]$, which corresponds to $C \approx 0.06$ for $n = 2$.
The ratios of the modified MRI characteristic scales to their standard counterparts are given by
\begin{align}\label{eq:ratios-2ndExample}
    \frac{\lambda_{\text{MRI}}}{\lambda_{\text{eMRI}}} &= \sqrt{\left|1 - \frac{1}{15}\left[\tilde C^2 (r - 1)^2 + 2\tilde C (r-1)\right]\right|}\;, \\
    \frac{\tau_{\text{MRI}}}{\tau_{\text{eMRI}}} &= \left|1 + \frac{1}{3}\tilde C (r - 1)\right|\;.
\end{align}
\begin{figure*}
    \centering
    \includegraphics[width=0.95\linewidth]{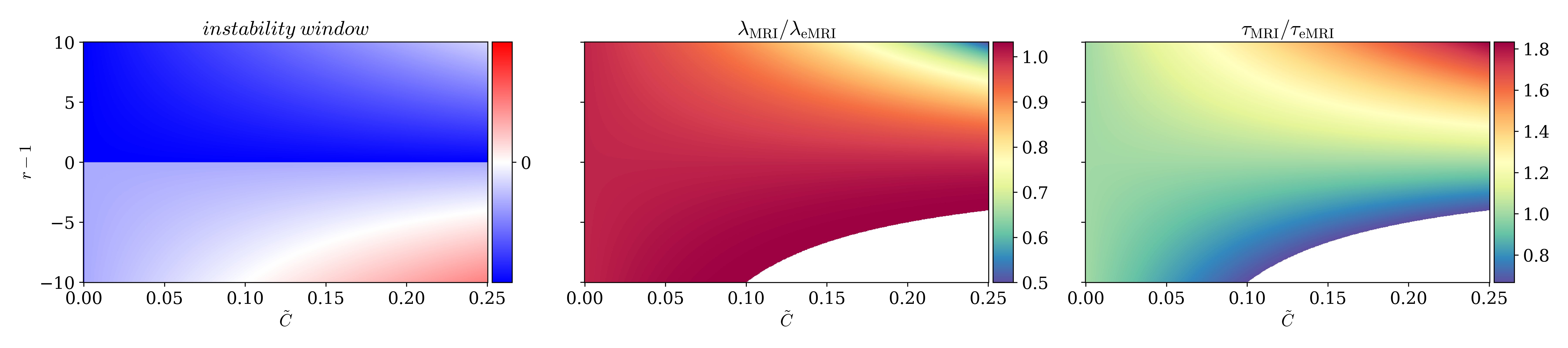}
    \caption{\textit{Extended MRI analysis for the analytical model given by a vertical and toroidal magnetic field.} \textit{From left to right}: Analytic instability window, ratio of the wavelength of the fastest-growing mode, and ratio of the timescale of the instability in the generalized case vs the standard MRI result. The wavelength and timescale of the fastest-growing mode are shown only where the instability is active according to the generalized criterion.}
    \label{fig:AnalyticInsta2D}
\end{figure*}

The instability window and the corresponding ratios of characteristic scales are shown in \cref{fig:AnalyticInsta2D}. The origin of each panel corresponds to the standard MRI, located in a regime where magnetic field gradients are small compared to the velocity shear. A key feature is the line $r = 1$, where the radial gradient scales of the magnetic field in the two components are equal. Along this line, the ratio of scales is exactly one, regardless of the magnitude of the gradients, as seen in \cref{eq:ratios-2ndExample}.
Deviations from unity in the plotted ratios remain modest for small values of $f$, and become more pronounced with stronger gradients and/or significant mismatches between $L_z$ and $L_y$. The impact of magnetic gradients on the instability window is particularly evident in the lower half-plane, where instability is suppressed at lower values of $\tilde{C}$, especially in the presence of strong mismatches. In contrast, the instability is never fully suppressed in the upper half-plane for the parameter values considered.
The timescale ratio also displays a clear asymmetry: the standard MRI timescale is larger than the extended one in the upper half-plane---indicating that gradients enhance the growth of unstable modes---while it is smaller in the lower half-plane, where gradients reduce the growth rate. As in the previous example, the reduction remains moderate, with deviations in the lower half-plane typically limited to $20$--$30\%$. 
Finally, the wavelength ratio is close to one across most of the parameter space explored. However, for sufficiently strong gradients and mismatches, the characteristic wavelength becomes noticeably shorter in the upper half-plane.

\section{Application to a numerical simulation of binary neutron star remnants}\label{sec:BNSdiagnostic}
In this section we investigate the impact of magnetic gradients on the generalized MRI phenomenology associated with a more extreme scenario: the long-lived, hyper-massive neutron star produced by a binary merger. The numerical solution was obtained by solving the Einstein equations coupled to the general relativistic MHD equations \citep{andersson2017beyond,andersson2022GRMHD} through the coalescence of a binary neutron star system. In particular, we considered the outcome of the simulations by 
\citet{TurbBampBNS_2022}, \citet{RoleTurbRicard2023} and \citet{DelayedJet24}, which were able to at least partially resolve the small-scale dynamics and lead to a converging saturation of the
magnetic energy, dominated by winding and small scales, starting from a much smaller magnetic field seed. These results were possible by a combination of high-resolution, high-order numerical schemes, and the use of advanced techniques to capture the sub-grid scales present during the turbulent regime. 
It is important to highlight that, while the analysis described in the previous sections was carried out within the non-relativistic framework, the simulation we considered was performed using full general relativity. 
This approach remains reasonable nonetheless, as the MRI is derived within the non-relativistic framework before being applied to the context of binary mergers. 

\subsection{Numerical set-up and analysis}
The numerical simulation was performed using the code {\sc MHDuet}, which is automatically generated by the platform {\sc Simflowny} \citep{arbona13,arbona18} to run on the {\sc SAMRAI} infrastructure \citep{hornung02,gunney16} with parallelization and adaptive mesh refinement. It uses fourth-order-accurate operators for the spatial derivatives in the Einstein equations, supplemented with sixth-order Kreiss-Oliger dissipation; a high-resolution shock-capturing method for the fluid, based on the  Lax-Friedrich flux splitting formula \citep{shu98} and the fifth-order reconstruction method MP5 \citep{suresh97}; a fourth-order Runge-Kutta scheme with sufficiently small time step $\Delta t \leq 0.4~\Delta x$ (where $\Delta x$ is the grid spacing); and an efficient and accurate treatment of the refinement boundaries when sub-cycling in  time~\citep{McCorquodale:2011,Mongwane:2015}. A complete assessment of the accuracy and robustness of the numerical methods implemented can be found in \cite{palenzuela18,vigano19}.
The binary neutron star is evolved in a cubic domain of size
$[-1228, 1228]$km using eight refinement levels, each with twice the resolution of the previous. The finest level covers the region where the density is $\rho \geq 5\times10^{12}$g  cm$^{-3}$. In this region, we achieve a maximum resolution of $\Delta x_{min} = 60$m during the first $110$ms after the merger, which is decreased to $\Delta x_{min} = 120$m later on.

For our post-process analysis, and due to quasi-axisymmetry of the solution at late times, it is more convenient to compute averages of the relevant fields along the azimuthal direction $\varphi$.
Therefore, we decomposed a generic field $U (R,z,\varphi)$, which can be either a scalar or a component of a tensor field, as the average $\bar U (R,z)$ plus a residual $\delta U ( R,z,\varphi)$. We represent 2D maps of these averaged fields as functions of the cylindrical coordinates $(R,z)$.
To be more specific, the averages were calculated using
\begin{align}\label{eq:azimuthal_average}
    \bar U (R,z) &= \frac{\int_0^{2\pi}U (R,\varphi,z) \chi^{-1/2} R d\varphi}{ \int_0^{2\pi} \chi^{-1/2}Rd\varphi} \;.
\end{align}
where the cylindrical coordinates are centred on the system centre of mass---the cylindrical axis $z$ is the same as the one relative to the Cartesian coordinates used throughout the simulation (i.e. perpendicular to the orbital plane)---and $\chi = \gamma^{-1/3}$ is the conformal factor of the spatial metric\footnote{We note that we computed the averages in this way as we were considering the simplifying assumption of a conformally flat spatial metric.} $\gamma_{ij}$ in the standard 3+1 decomposition \citep{mizuno2025general,palenzuela2020NR,gourgoulhon20123+1}. 
These azimuthal averages are performed considering only the high-density regions with $\rho \geq 5\times10^{10}$~g~ cm$^{-3}$, which contain the bulk of the remnant mass and are less sensitive to artefacts arising from the low inertia treatment (simulations use a floor value $6 \times10^{4}$~g~ cm$^{-3}$ for density). 

\begin{figure*}
    \centering
    \includegraphics[width=0.9\linewidth]{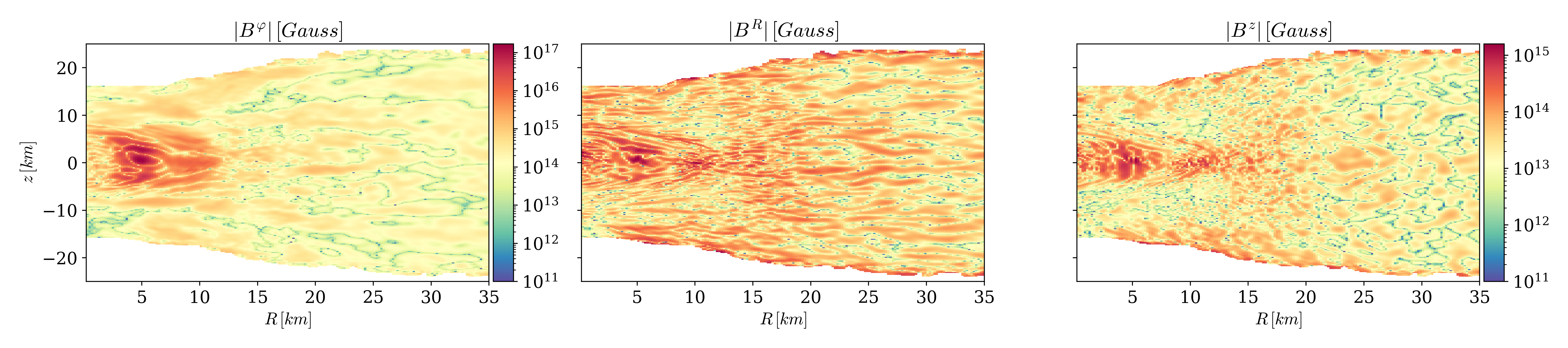}
    \caption{\textit{2D maps of the azimuthally averaged magnetic field magnitudes at $t\simeq 40$~ms after merger.} \textit{From left to right}: 2D maps of the magnitude of the radial, azimuthal, and vertical component of the magnetic field. We see the different components present large gradients, which we take into account in the generalized MRI criterion.}
    \label{fig:2Dmaps_Bintensity_t40}
\end{figure*}

In \cref{fig:2Dmaps_Bintensity_t40} we plot the 2D maps of the magnitude of the various magnetic field components at $t=40$~ms.
In particular, we observe that the magnetic field components present large gradients, which suggests that they need to be considered to study the conditions for MRI. These gradients are in both vertical and radial directions and particularly evident for the azimuthal and vertical components. The variations seem more ordered close to the orbital plane, while they are more stochastic in the less dense regions, where also the azimuthal variations (averaged out by definition) are more important (see Appendix~\ref{app:SimContact} for a detailed analysis).
In \cref{fig:InstaMasking} we show the angular velocity (left panel) and its (logarithmic) derivative (middle panel). As it is expected in hyper-massive neutron star remnants, the angular velocity decays both with the cylindrical radii and with the absolute value of the z coordinate (i.e. as we move away perpendicularly from the equatorial plane).

\subsection{Comparison of the standard versus extended MRI analysis}

\begin{figure*}
    \centering
    \includegraphics[width=0.9\linewidth]{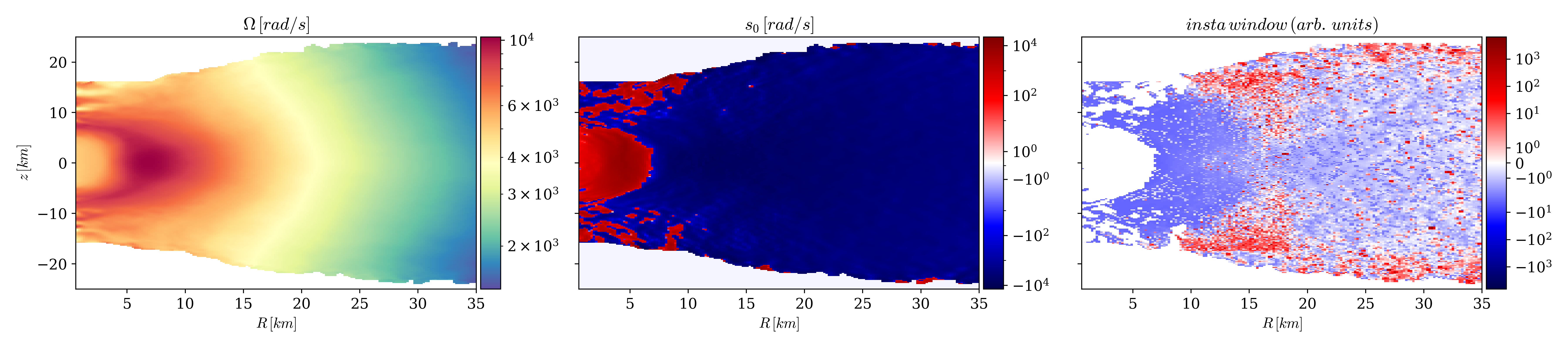}
    \caption{\textit{Instability window calculation at $t\simeq40$~ms after merger}. From the angular frequency ($\Omega$) we computed its gradient ($s_0$) and mask those points where $s_0>0$, namely those for which the standard MRI would be inactive. Then we evaluated the updated instability criterion, considering at once the two cases $f>0$ and $f<0$. In the right panel,  blue points are unstable according to the updated MRI criterion, and red are those for which the magnetic gradients are large enough to switch off the instability; all coloured points are MRI-unstable according to the standard criterion.}
    \label{fig:InstaMasking}
\end{figure*}

\begin{figure}
    \centering
    \includegraphics[width=0.35\textwidth]{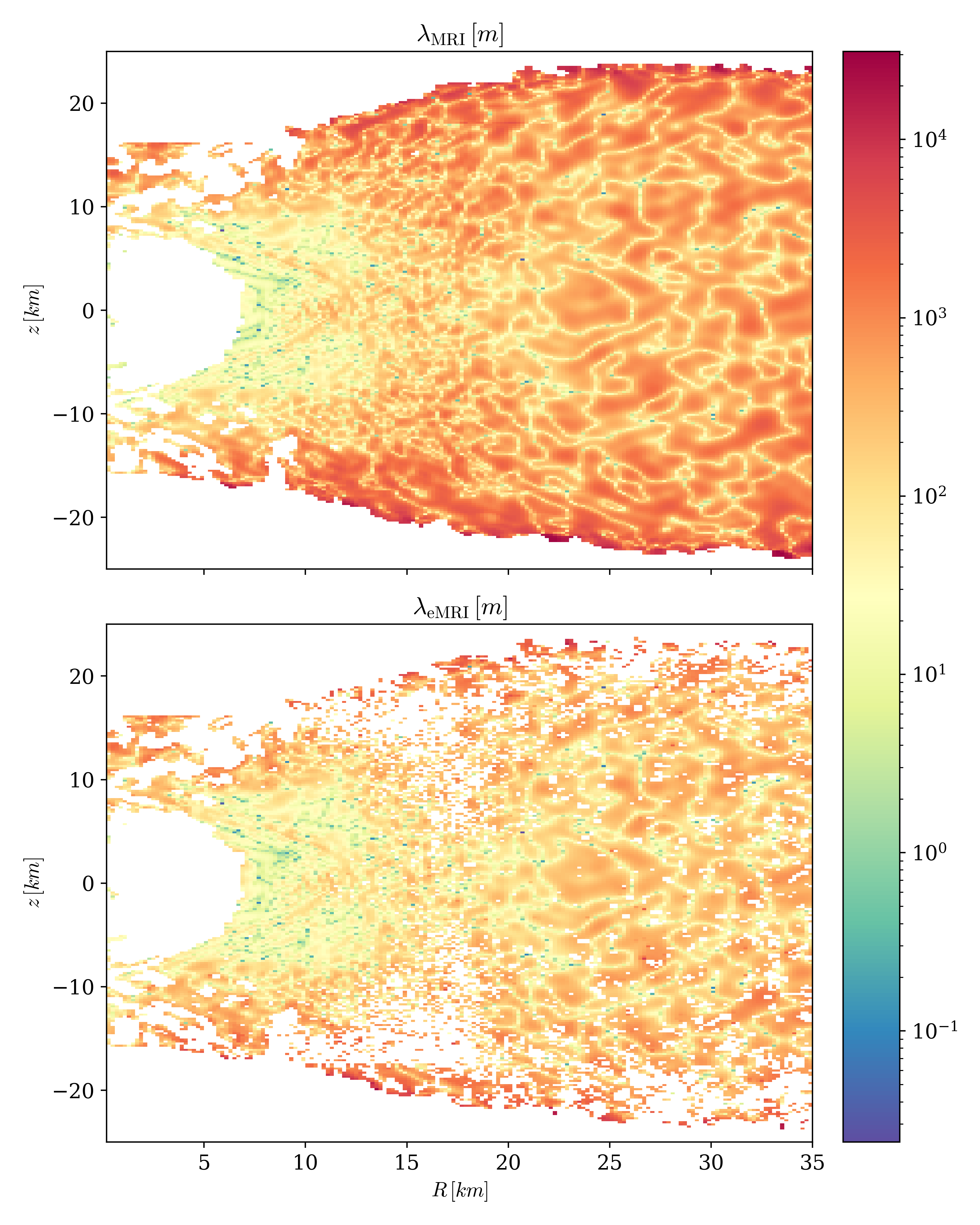}
    \caption{\textit{Comparison of the wavelength of the fastest-growing mode in the standard vs generalized case, at $t\simeq40$~ms after merger}. \textit{Top}: Standard MRI case, all coloured points are unstable to the MRI. \textit{Bottom}: Generalized case. We mask the points where the instability condition is not met. Where the generalized instability is active, we observe $\lambda_{\text{eMRI}}$ to reach values close to but slightly smaller than the standard case. Most interestingly, $\lambda_{max}$ is of the order of $100$~m or smaller at small radii, $r\lesssim 12$~km.}
    \label{fig:LambdaComparison40}
\end{figure}

\begin{figure}
    \centering
    \includegraphics[width=0.35\textwidth]{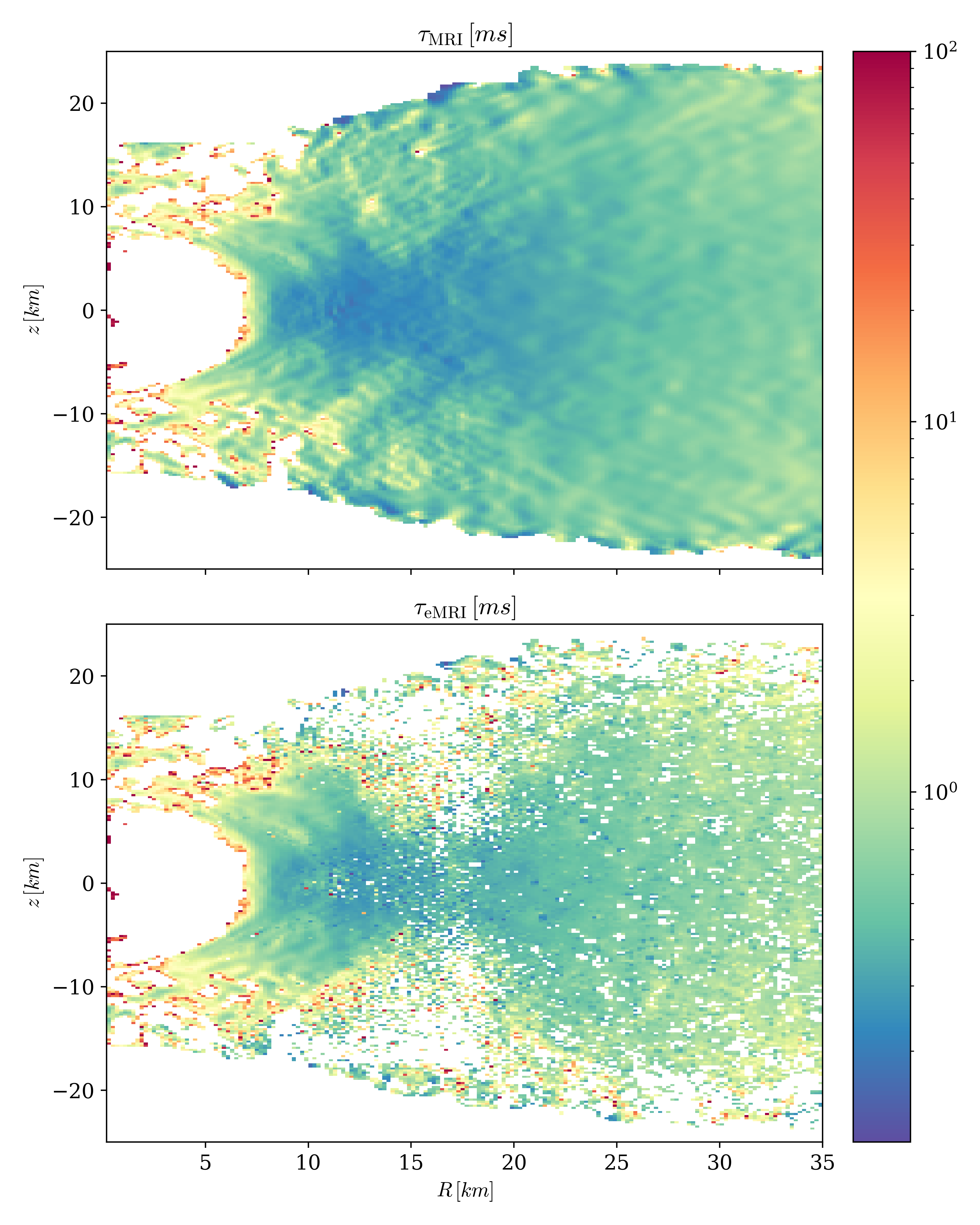}
    \caption{\textit{Comparison of the timescale of the fastest-growing mode in the standard vs generalized case, at $t\simeq40$ms after merger}. \textit{Top}: Standard MRI case. All coloured points are unstable to the MRI, and we observe the $\tau_{\text{MRI}}$ is a fraction of a millisecond at almost all radii $\lesssim 30$km, thus making the instability sufficiently fast to impact the post-merger dynamics. \textit{Bottom}: Updated case. We mask the points where the instability condition is not met. Where the generalized instability is active, we observe that $\tau_{\text{eMRI}}$ is  close to or slightly larger than $1$ms, particularly at large radii, $r\gtrsim15$km. This means that the instability takes longer to develop according to the generalized criterion, and might be too slow to significantly impact the post-merger dynamics.}
    \label{fig:TauComparison40}
\end{figure}

Here we describe in detail the procedure for calculating a new instability window using the extended MRI criteria, as well as the predictions for $\lambda_{eMRI}$ and $\tau_{eMRI}$. The required steps are represented in \cref{fig:InstaMasking} for an illustrative time snapshot at $t=40$ms. 
First, using the angular velocity we could easily compute $S_0$ (cf. \cref{eq:corotatingObserver}) and we mask those points where $S_0>0$, namely those for which the standard MRI would be inactive. Then we evaluated the updated instability criterion, considering at once the two cases $f>0$ and $f<0$ (cf. \cref{eMRIcriterion_fneg,eMRIcriterion_fpos}). The points in the right panel that are blue are unstable according to the new generalized MRI criterion, whereas the red ones are those for which the magnetic gradients are large enough to shut down the instability. All coloured points in this panel would be MRI-unstable according to the standard criterion.
Notice that including the new branch of the instability $S_0\geq 0$ has no impact at all in this plot, since our solution never satisfies the quite restrictive conditions
$\kappa^2 - 4\Om^2 \leq |f| \leq \kappa^2$. 
 
We next compared the predictions for $\lambda_{eMRI}$ and $\lambda_{MRI}$, displayed in \cref{fig:LambdaComparison40}.
In the standard MRI case (upper plot), we observe that $\lambda_{\text{MRI}}$ can reach values of $10^3-10^4$~m, especially at large radii, $R\gtrsim 15$~km, while it is considerably smaller, $\lambda_{\text{MRI}}\sim 10$~m, at $R\lesssim 15$~km. The extended MRI case is considered in the lower panel. First of all, one can note that the potentially unstable regions are reduced compared to the standard criterion. The regions excluded by the criteria extension are mainly the ones having the largest $\lambda_{\rm MRI}$. The potentially unstable region is mainly composed by a pseudo-torus, with a maximum $\sim 20$km in height and extended from $\sim 7$ to $\sim 17$km in the orbital plane. In these regions, $\lambda_{\text{eMRI}}$ reaches values comparable to the standard case, $\lambda_{\text{MRI}}$, being of the order of $100\text{m}$ or smaller at small radii $R\lesssim 12\text{km}$. Formally, there are extended potential MRI-unstable regions, at larger $R$ and $z$, but their $\lambda_{\rm eMRI}$ is larger than in the main pseudo-torus. A wavelength that is too large may preclude MRI, as the linear analysis is valid only when $\lambda_{\rm eMRI}$ is much smaller than the characteristic length-scales of velocity and magnetic field variations.

Finally, we could compare the values of $\tau_{eMRI}$ and $\tau_{MRI}$, which are displayed in \cref{fig:TauComparison40}.
The $\tau_{\text{MRI}}$ obtained from the standard MRI is a fraction of a millisecond for most of the region $R\lesssim 30$km, which makes the instability sufficiently fast to have an impact on the post-merger dynamics. The one calculated with the generalized MRI analysis is displayed in the lower panel, where we mask the points in which the instability condition is not satisfied. When the generalized instability is active, we observe that $\tau_{\text{eMRI}}$ is close to or slightly larger than $1 \text{ms}$, particularly at large radii $R\gtrsim 15$~km.
This is important, because such timescales are comparable to the dynamical timescales of the remnant. In other words, the basic assumption that the velocity and the large-scale (background) field change much more slowly than the instability growth time is not guaranteed, which might further hamper the possibility of developing MRI.

\subsection{Evolution of the MRI quantities}
\begin{figure*}
    \centering
    \includegraphics[width=0.9\linewidth]{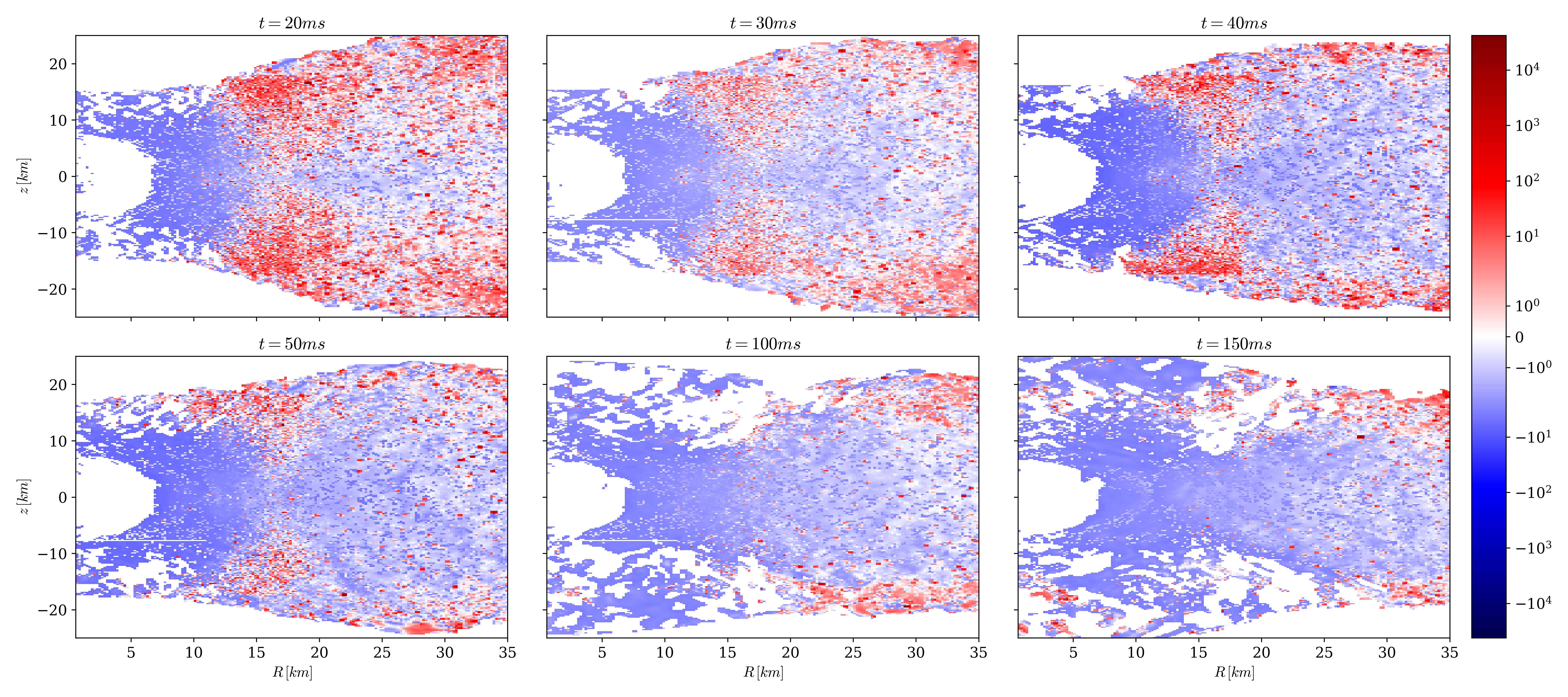}
    \caption{\textit{Evolution of the generalized 2D instability window}. All coloured points are MRI-unstable according to the standard MRI criterion, while only the blue ones are according to the updated criterion. We see that at $20$ms the instability is potentially active only at small radii, $7 \lesssim R \lesssim 12$km, and close to the equatorial plane. As time evolves, we observe that the region at larger radii, but still close to the equatorial plane, also becomes more and more suited for the instability to develop, e.g. at $t\simeq 50$ms after merger.}
    \label{fig:Insta2DEvolution}
\end{figure*}

So far, we have illustrated the analysis at a given, representative time. In order to see the evolution, we show how the instability window, wavelength, and growth timescale evolve in time, in \cref{fig:Insta2DEvolution}. We see that at $t \simeq 20~\text{ms}$ the instability is possible only at small radii $7 \lesssim R \lesssim 12$~km, and close to the equatorial plane. As time evolves (e.g. at $t \gtrsim 50$~ms after merger), the region extends to larger radii, but still close to the orbital plane, becomes more and more suited for the instability to develop.
Away from these preferred regions, there are other locations in the $R-z$ plane where MRI could potentially be active, although these appear to be somewhat spatially disconnected. For a region to be susceptible to MRI activity, it must be sufficiently extended to accommodate multiple wavelengths of the unstable modes.

\begin{figure*}
    \centering
    \includegraphics[width=0.9\linewidth]{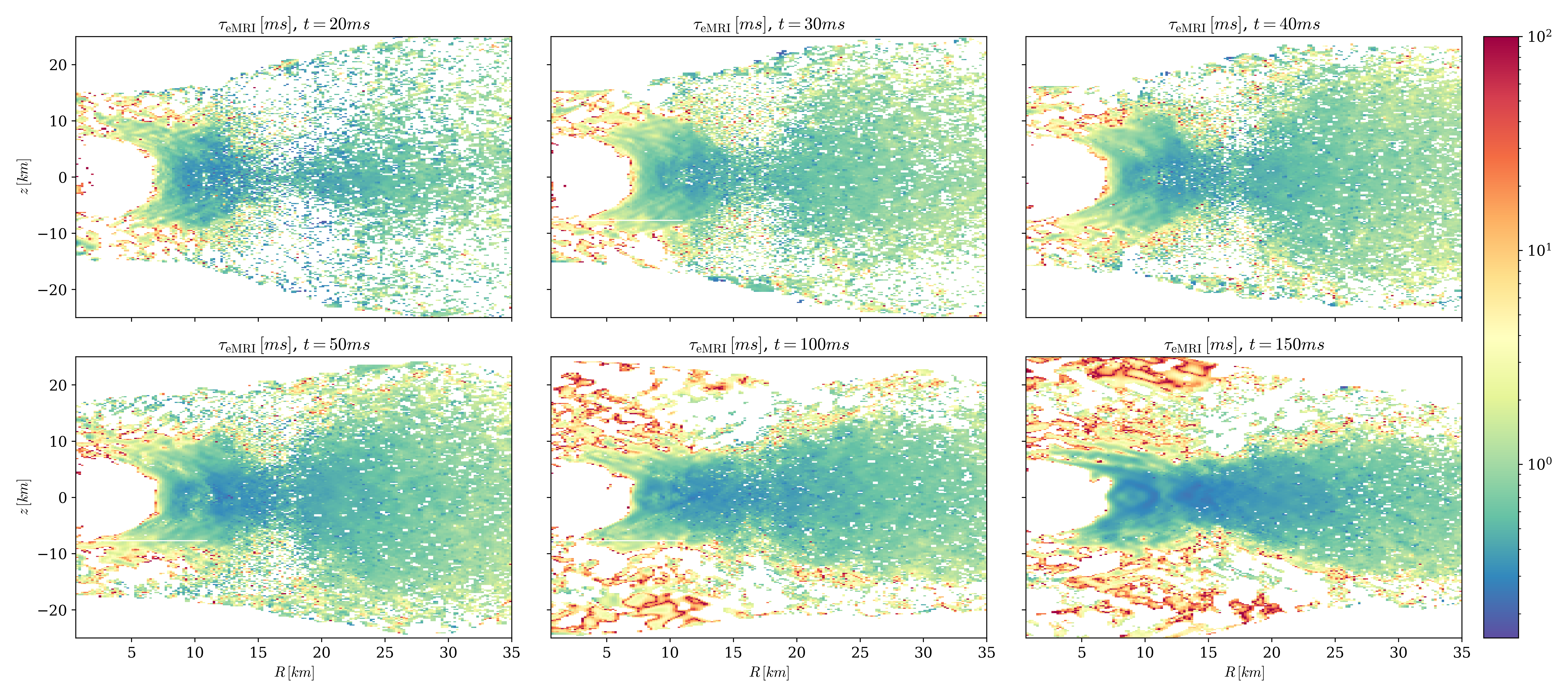}
    \caption{\textit{Evolution of the timescale of the fastest-growing mode, $\tau_{\text{eMRI}}$, of the extended MRI}. All coloured points meet the generalized instability criteria. }
    \label{fig:TauMin2DEvolution}
\end{figure*}

The evolution of the modified $\tau_{\text{eMRI}}$ is displayed in \cref{fig:TauMin2DEvolution}. Focusing on the region close to the equatorial plane---the one that becomes progressively more suited for the instability to develop---we observe that the timescale of the instability decreases steadily, and is of the order of a fraction of a millisecond at late times. This means that the central region becomes more prone to the instability, but also that the instability would grow on shorter timescales. This suggests that the MRI only acquires an important role at late times $t \simeq  100 \text{ms}$.

\begin{figure*}
    \centering
    \includegraphics[width=0.9\linewidth]{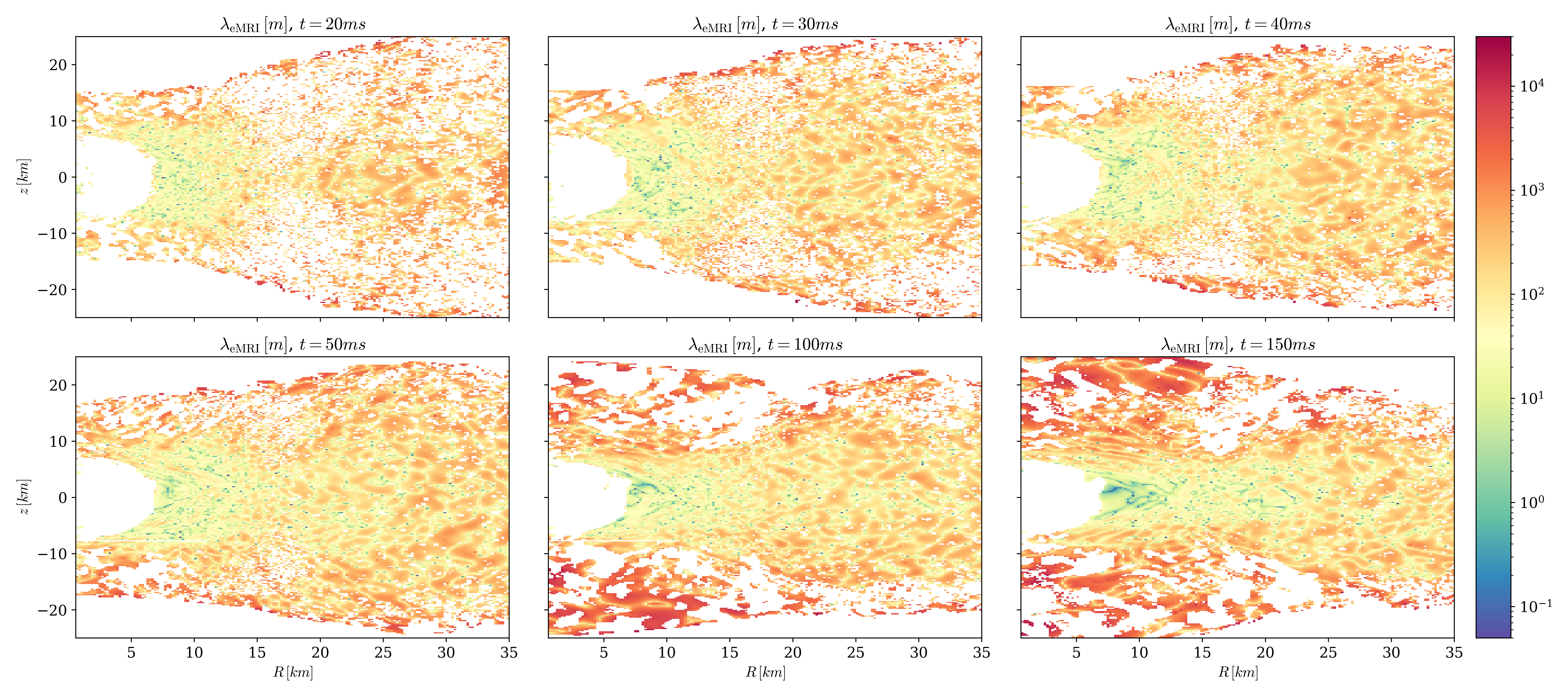}
    \caption{\textit{Evolution of the wavelength of the fastest-growing mode,  $\lambda_{\text{eMRI}}$, of the extended MRI}. All coloured points meet the generalized instability criteria. Focusing on the region close to the equatorial plane---the one that becomes progressively more suited for the instability to develop---we see that $\lambda_{\text{eMRI}}$ becomes progressively smaller and smaller, of the order of $1-10$m at late times. This means that in the central region the characteristic wavelength of the instability is small, thus making the instability more difficult to resolve.}
    \label{fig:LambdaMax2DEvolution}
\end{figure*}

The evolution of $\lambda_{\text{eMRI}}$ is plotted in \cref{fig:LambdaMax2DEvolution}. Focusing again in the region close to the equatorial plane, we can observe that $\lambda_{\text{eMRI}}$ decreases progressively,  reaching values of the order of $1-10 \text{m}$ at late times. Consistently with what we just said about the decreasing growth time, this is promising since $\lambda_{\rm eMRI}$ is much smaller than the magnetic and velocity field variation length-scales. However, numerically, it poses hard challenges to resolve it, due to an increasingly demanding spatial resolution over a larger region.

\begin{figure}
    \centering
    \includegraphics[width=0.9\linewidth]{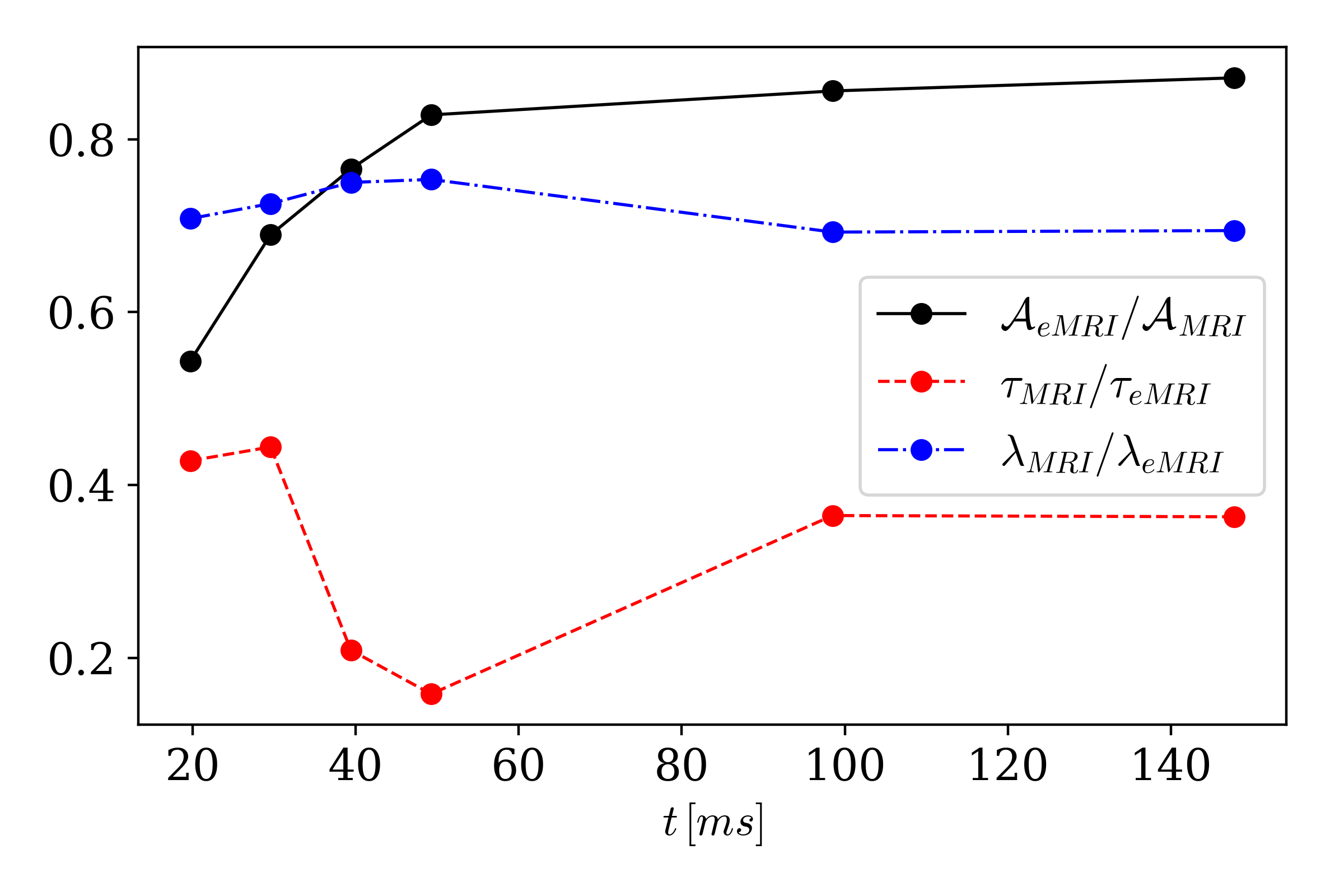}
    \caption{\textit{Averaged ratios of relevant quantities as a function of time}. We show the time evolution of three ratios: the area in which the extended MRI is active ($\mathcal{A}_{eMRI}$) over the area in which the standard MRI is active ($\mathcal{A}_{MRI}$); the ratio $\lambda_{MRI}/\lambda_{eMRI}$; and the ratio $\tau_{MRI}/\tau_{eMRI}$. Interestingly, the low value of $\tau_{MRI}/\tau_{eMRI}$ (especially when multiplied by $\mathcal{A}_{MRI}/\mathcal{A}_{eMRI}$ to estimate the effective ratio) indicates that the magnetic field gradients might be diminishing the MRI.}
    \label{fig:average_MRI_time}
\end{figure}

All these qualitative results can be summarized quantitatively in \cref{fig:average_MRI_time},
where the following averaged ratios over the 2D maps are represented as a function of time: the area in which the extended MRI is active ($\mathcal{A}_{eMRI}$) over the area in which the standard MRI is active ($\mathcal{A}_{MRI}$); the ratio $\lambda_{MRI}/\lambda_{eMRI}$; and the ratio $\tau_{MRI}/\tau_{eMRI}$. As we discussed previously, the instability region with the generalized criterion is smaller than with the standard one. 
The ratio $\lambda_{MRI}/\lambda_{eMRI}$ is close to one because the values are similar when both instability criteria (standard and extended) are satisfied. 
As we noticed earlier though, $\lambda_{MRI}$ takes on the largest values only when the standard MRI is active. 
This means that, effectively, it is harder to capture numerically the MRI according to the extended criterion.
Even most relevant is the result of the ratios $\tau_{MRI}/\tau_{eMRI}$, which shows that the growth rate is significantly shorter when magnetic field gradients are considered in the analysis. One can include also the ratio of the areas to estimate an effective ratios $\tau_{MRI}/\tau_{eMRI}$, which is even smaller than the previous, especially at early times. These results suggest that the magnetic field gradients might be suppressing (or at least slowing down) the axisymmetric MRI.

\section{Outlook and conclusions}\label{sec:conclusions}
The MRI is considered a key mechanism to induce turbulence in accretion disks and is believed to be a key process mediating the emergence of large-scale, poloidal magnetic fields in the post-merger remnants of compact binary mergers \citep{duez2006collapse,siegel2013MRI,shibata2015numerical,ruiz2016BNSjet,ciolfi2019_100msBNS,Ciolfi2020_BinBNS,radice2024turbulence,ETbluebook}. 
These simulations show a complex magnetic field topology, which differs significantly from the simplified conditions under which MRI was originally formulated.
In particular, certain simulations of long-lived remnants fail to provide evidence of poloidal magnetic field amplification after the vigorous Kelvin-Helmholtz amplification phase---while clearly showing the impact of linear winding in the toroidal component \citep{TurbBampBNS_2022,RoleTurbRicard2023,DelayedJet24}. 

In this study we tackled this question by investigating, for the first time, how complex magnetic field topologies, specifically magnetic field gradients, influence the axisymmetric MRI. 
The strategy adopted here is parallelled by the analysis of \citet{CeloraMRI}, who show that the conventional MRI criteria and results crucially depend on the assumption of an axisymmetric background.
After having explored the impact of several magnetic field background configurations, we find that radial magnetic field gradients can have a notable impact on the MRI: the `instability window' is reduced and both the wavelength and growth rate of the MRI's fastest-growing mode are altered.

Next, we examined the generalized conditions for the MRI using two simplified models inspired from the accretion and protoplanetary disks described in the literature \citep{pelletier1992hydromagnetic,galli06,jafari19}. 
This exercise serves a dual purpose: firstly, it demonstrates that while our initial focus is on binary neutron star applications, our findings have broader implications. 
Secondly, it helps in developing an intuition for how additional magnetic gradients influence the instability.

Finally, we investigated the generalized MRI conditions in the post-merger environment by postprocessing a simulation of a long-lived binary neutron star merger remnant. 
In particular, the simulation considered effectively captures magnetic field growth from an initial realistic field strength of approximately $10^{11}$ G \citep{palenzuela18,vigano19}.
Our findings indicate that the (generalized) MRI can be active primarily in regions of small radii ($5\lesssim R\lesssim 15$km) during the early stages ($t \lesssim 40$ms). 
As time progresses, the region close to the equatorial plane ($-10\lesssim z\lesssim10$km) and large radii ($R\gtrsim 15$km) also become susceptible to MRI development. 
However, the e-folding timescale of the instability remains consistently close to or slightly above $1$ms across all radii and up to $t\gtrsim 50$ms after mergers. 
This timescale decreases to a fraction of a millisecond (the typical value according to conventional MRI) only at late times, $t \gtrsim 100$ms. Additionally, we observed that the wavelength of the fastest-growing MRI mode is generally shorter compared to the standard case: of the order of $10-100$ m at most in regions where the MRI would potentially be active, namely small radii at early times, or all radii but close to the equatorial plane at late times. 
Notably, the value at small radii is consistently around $1-10$ m throughout the simulation, particularly at late times.
Such small values are, on the one hand, promising for the applicability of linear analysis, as a key requirement for its validity is that the dominant unstable MRI modes have wavelengths much shorter than the characteristic scales over which the magnetic and velocity fields vary. On the other hand, accurately resolving such short wavelengths presents significant challenges for current computational resources---see also the discussion in \cref{app:QualityFactors}, where we report on the empirical quality criteria commonly used to quantify the ability of a simulation to capture MRI-driven MHD turbulence.
A final caveat we feel compelled to raise concerns the magnetic non-axisymmetric indicators discussed in \cref{app:SimContact}, which remain close to unity across much of the domain even at late times, $\mathcal{O}(100)$ ms. While we caution that non-axisymmetric indicators being close to unity does not necessarily imply the absence of growth of an axisymmetric large-scale poloidal field, this nevertheless further challenges the expected influence of the axisymmetric MRI. It also suggests that the characteristic timescales and length-scales of magnetic field variations are significantly shorter than commonly assumed.
This caveat could also apply to certain protoplanetary disks where no organized large-scale magnetic field is observed \citep{huang24,huang25}.

In summary, the results presented in this paper offer a coherent explanation for the lack of growth in the poloidal component (after the Kelvin-Helmholtz unstable phase) in the simulation examined here. 
Most notably, the longer timescales associated with the generalized instability, combined with its later onset, would suggest that the axisymmetric MRI's influence on post-merger dynamics is significantly reduced. 
In turn, this indicates that the role of the MRI in affecting the post-merger dynamics should be approached with care in modelling efforts. 
Finally, we caution that setting an unrealistically high initial large-scale magnetic field, instead of starting with a weak field amplified by KHI and winding, biases the post-merger configuration, and the magnetic gradients in particular {\citep{DelayedJet24,2025arXiv250618995G}.

\begin{acknowledgements}
TC is an ICE Fellow and is supported through the Spanish program Unidad de Excelencia Maria de Maeztu CEX2020-001058-M. CP's work was supported by the Grant PID2022-138963NB-I00 funded by the Spanish Ministry of Science and Innovation, MCIN/AEI/10.13039/501100011033/FEDER, UE. DV is supported through the Spanish program Unidad de Excelencia Maria de Maeztu CEX2020-001058-M as well as by the European Research Council (ERC) under the European Union’s Horizon 2020 research and innovation program (ERC Starting Grant ”IMAGINE” No. 948582, PI: DV). RA-M is funded by the Deutsche Forschungsgemeinschaft (DFG, German Research Foundation) under the Germany Excellence Strategy - EXC 2121 'Quantum Universe' - 390833306. The authors thankfully acknowledge RES resources provided by BSC in MareNostrum, projects AECT-2024-1-0010, AECT-2024-2-0004 and AECT-2024-3-0007.
\end{acknowledgements}

\bibliographystyle{aa}
\bibliography{biblio}    

\begin{thebibliography}{78}
\expandafter\ifx\csname natexlab\endcsname\relax\def\natexlab#1{#1}\fi

\bibitem[{Abac {et~al.}(2025)}]{ETbluebook}
Abac, A. {et~al.} 2025, to appear in JCAP [\eprint[arXiv]{2503.12263}]

\bibitem[{Aguilera-Miret {et~al.}(2024)Aguilera-Miret, Palenzuela, Carrasco, Rosswog, \& Vigan\`o}]{DelayedJet24}
Aguilera-Miret, R., Palenzuela, C., Carrasco, F., Rosswog, S., \& Vigan\`o, D. 2024, Phys. Rev. D, 110, 083014

\bibitem[{Aguilera-Miret {et~al.}(2023)Aguilera-Miret, Palenzuela, Carrasco, \& Vigan\`o}]{RoleTurbRicard2023}
Aguilera-Miret, R., Palenzuela, C., Carrasco, F., \& Vigan\`o, D. 2023, Phys. Rev. D, 108, 103001

\bibitem[{Anderson {et~al.}(2008)Anderson, Hirschmann, Lehner, Liebling, Motl, Neilsen, Palenzuela, \& Tohline}]{anderson2008magnetized}
Anderson, M., Hirschmann, E.~W., Lehner, L., {et~al.} 2008, Physical Review Letters, 100, 191101

\bibitem[{Andersson {et~al.}(2022)Andersson, Hawke, Celora, \& Comer}]{andersson2022GRMHD}
Andersson, N., Hawke, I., Celora, T., \& Comer, G. 2022, Monthly Notices of the Royal Astronomical Society, 509, 3737

\bibitem[{Andersson {et~al.}(2017)Andersson, Hawke, Dionysopoulou, \& Comer}]{andersson2017beyond}
Andersson, N., Hawke, I., Dionysopoulou, K., \& Comer, G. 2017, Classical and Quantum Gravity, 34, 125003

\bibitem[{Anile(1989)}]{Anile}
Anile, A.~M. 1989, Relativistic fluids and magneto-fluids (Cambridge University Press)

\bibitem[{Arbona {et~al.}(2013)Arbona, Artigues, Bona-Casas, Massó, Miñano, Rigo, Trias, \& Bona}]{arbona13}
Arbona, A., Artigues, A., Bona-Casas, C., {et~al.} 2013, Comput. Phys. Commun., 184, 2321

\bibitem[{Arbona {et~al.}(2018)Arbona, Miñano, Rigo, Bona, Palenzuela, Artigues, Bona-Casas, \& Massó}]{arbona18}
Arbona, A., Miñano, B., Rigo, A., {et~al.} 2018, Comput. Phys. Commun., 229, 170

\bibitem[{{Balbus} \& {Hawley}(1991)}]{BalbusHawley1}
{Balbus}, S.~A. \& {Hawley}, J.~F. 1991, Astrophys. J., 376, 214

\bibitem[{{Balbus} \& {Hawley}(1992)}]{BalbusHawley4}
{Balbus}, S.~A. \& {Hawley}, J.~F. 1992, Astrophys. J., 400, 610

\bibitem[{Balbus \& Hawley(1998)}]{BalbusHawleyRev1998}
Balbus, S.~A. \& Hawley, J.~F. 1998, Rev. Mod. Phys., 70, 1

\bibitem[{Barletta(2022)}]{barletta2022boussinesq}
Barletta, A. 2022, Mechanics Research Communications, 124, 103939

\bibitem[{{Beltr{\'a}n} {et~al.}(2019){Beltr{\'a}n}, {Padovani}, {Girart}, {Galli}, {Cesaroni}, {Paladino}, {Anglada}, {Estalella}, {Osorio}, {Rao}, {S{\'a}nchez-Monge}, \& {Zhang}}]{beltran19}
{Beltr{\'a}n}, M.~T., {Padovani}, M., {Girart}, J.~M., {et~al.} 2019, \aap, 630, A54

\bibitem[{Carrasco {et~al.}(2020)Carrasco, Vigan{\`o}, \& Palenzuela}]{carrasco2020gradient}
Carrasco, F., Vigan{\`o}, D., \& Palenzuela, C. 2020, Physical Review D, 101, 063003

\bibitem[{Celora {et~al.}(2021)Celora, Andersson, Hawke, \& Comer}]{CeloraFibrLES}
Celora, T., Andersson, N., Hawke, I., \& Comer, G.~L. 2021, Phys. Rev. D, 104, 084090

\bibitem[{Celora {et~al.}(2024{\natexlab{a}})Celora, Andersson, Hawke, Comer, \& Hatton}]{CeloraHigherLevel}
Celora, T., Andersson, N., Hawke, I., Comer, G.~L., \& Hatton, M.~J. 2024{\natexlab{a}}, Phys. Rev. D, 110, 123039

\bibitem[{Celora {et~al.}(2024{\natexlab{b}})Celora, Hatton, Hawke, \& Andersson}]{CeloraLagrangFilter}
Celora, T., Hatton, M.~J., Hawke, I., \& Andersson, N. 2024{\natexlab{b}}, Phys. Rev. D, 110, 123040

\bibitem[{Celora {et~al.}(2023)Celora, Hawke, Andersson, \& Comer}]{CeloraMRI}
Celora, T., Hawke, I., Andersson, N., \& Comer, G.~L. 2023, Mon. Not. Roy. Astron. Soc., 527, 2437

\bibitem[{{Chandrasekhar}(1960)}]{Chandrasekar1960}
{Chandrasekhar}, S. 1960, Proc. Natl. Acad. Sci., 46, 253

\bibitem[{Ciolfi(2020)}]{Ciolfi2020_BinBNS}
Ciolfi, R. 2020, Gen. Rel. Grav., 52, 59

\bibitem[{Ciolfi {et~al.}(2019)Ciolfi, Kastaun, Kalinani, \& Giacomazzo}]{ciolfi2019_100msBNS}
Ciolfi, R., Kastaun, W., Kalinani, J.~V., \& Giacomazzo, B. 2019, Physical Review D, 100, 023005

\bibitem[{Combi \& Siegel(2023)}]{CombiSiegel2023}
Combi, L. \& Siegel, D.~M. 2023, Phys. Rev. Lett., 131, 231402

\bibitem[{Do{\u{g}}an \& Pek{\"u}nl{\"u}(2012)}]{dougan2012mri}
Do{\u{g}}an, S. \& Pek{\"u}nl{\"u}, E. 2012, Publications of the Astronomical Society of the Pacific, 124, 922

\bibitem[{Duez {et~al.}(2006)Duez, Liu, Shapiro, Shibata, \& Stephens}]{duez2006collapse}
Duez, M.~D., Liu, Y.~T., Shapiro, S.~L., Shibata, M., \& Stephens, B.~C. 2006, Physical Review Letters, 96, 031101

\bibitem[{Fromang \& Papaloizou(2007)}]{FromangZNF2007}
Fromang, S. \& Papaloizou, J. 2007, Astron. Astrophys., 476, 1113

\bibitem[{{Galli} {et~al.}(2006){Galli}, {Lizano}, {Shu}, \& {Allen}}]{galli06}
{Galli}, D., {Lizano}, S., {Shu}, F.~H., \& {Allen}, A. 2006, \apj, 647, 374

\bibitem[{Gogichaishvili {et~al.}(2017)Gogichaishvili, Mamatsashvili, Horton, Chagelishvili, \& Bodo}]{Gogichaishvili2017}
Gogichaishvili, D., Mamatsashvili, G., Horton, W., Chagelishvili, G., \& Bodo, G. 2017, Astrophys. J., 845, 70

\bibitem[{{Goodman} \& {Xu}(1994)}]{Goodman1994}
{Goodman}, J. \& {Xu}, G. 1994, \apj, 432, 213

\bibitem[{Gourgoulhon(2012)}]{gourgoulhon20123+1}
Gourgoulhon, E. 2012, 3+ 1 formalism in general relativity: bases of numerical relativity, Vol. 846 (Springer Science \& Business Media)

\bibitem[{Gunney \& Anderson(2016)}]{gunney16}
Gunney, B. T.~N. \& Anderson, R.~W. 2016, J. Parallel. Distr. Com., 89, 65

\bibitem[{{Guti{\'e}rrez} {et~al.}(2025){Guti{\'e}rrez}, {Cook}, {Radice}, {Bernuzzi}, {Fields}, {Hammond}, {Daszuta}, {Bandyopadhyay}, \& {Jacobi}}]{2025arXiv250618995G}
{Guti{\'e}rrez}, E.~M., {Cook}, W., {Radice}, D., {et~al.} 2025, arXiv e-prints, arXiv:2506.18995

\bibitem[{Hawley {et~al.}(2011)Hawley, Guan, \& Krolik}]{HawleyConvergence2011}
Hawley, J.~F., Guan, X., \& Krolik, J.~H. 2011, Astrophys. J., 738, 84

\bibitem[{Hawley {et~al.}(2013)Hawley, Richers, Guan, \& Krolik}]{HawleyConvergence2013}
Hawley, J.~F., Richers, S.~A., Guan, X., \& Krolik, J.~H. 2013, Astrophys. J., 772, 102

\bibitem[{Held \& Mamatsashvili(2022)}]{Held2022}
Held, L.~E. \& Mamatsashvili, G. 2022, Mon. Not. Roy. Astron. Soc., 517, 2309

\bibitem[{Herault {et~al.}(2011)Herault, Rincon, Cossu, Lesur, Ogilvie, \& Longaretti}]{Herault2011}
Herault, J., Rincon, F., Cossu, C., {et~al.} 2011, Phys. Rev. A, 84, 036321

\bibitem[{Hornung \& Kohn(2002)}]{hornung02}
Hornung, R. \& Kohn, S. 2002, Concurr. Comp. Pract. E., 14, 347

\bibitem[{{Huang} {et~al.}(2024){Huang}, {Girart}, {Stephens}, {Fern{\'a}ndez L{\'o}pez}, {Arce}, {Carpenter}, {Cortes}, {Cox}, {Friesen}, {Le Gouellec}, {Hull}, {Karnath}, {Kwon}, {Li}, {Looney}, {Megeath}, {Myers}, {Murillo}, {Pineda}, {Sadavoy}, {S{\'a}nchez-Monge}, {Sanhueza}, {Tobin}, {Zhang}, {Jackson}, \& {Segura-Cox}}]{huang24}
{Huang}, B., {Girart}, J.~M., {Stephens}, I.~W., {et~al.} 2024, \apjl, 963, L31

\bibitem[{{Huang} {et~al.}(2025){Huang}, {Girart}, {Stephens}, {Myers}, {Zhang}, {Cortes}, {S{\'a}nchez-Monge}, {Fern{\'a}ndez L{\'o}pez}, {Le Gouellec}, {Megeath}, {Murillo}, {Carpenter}, {Li}, {Liu}, {Looney}, {Sadavoy}, {Karnath}, \& {Kwon}}]{huang25}
{Huang}, B., {Girart}, J.~M., {Stephens}, I.~W., {et~al.} 2025, \apj, 984, 29

\bibitem[{{Jafari}(2019)}]{jafari19}
{Jafari}, A. 2019, arXiv preprint:1904.09677

\bibitem[{Jiang {et~al.}(2025)Jiang, Ng, Chabanov, \& Rezzolla}]{jiang2025LongTermBImpact}
Jiang, J.-L., Ng, H. H.-Y., Chabanov, M., \& Rezzolla, L. 2025, Phys. Rev. D, 111, 103043

\bibitem[{Kalinani {et~al.}(2025)Kalinani, Ciolfi, Campanelli, Giacomazzo, Pavan, Wen, \& Zlochower}]{KalinaniJetNoatmo}
Kalinani, J.~V., Ciolfi, R., Campanelli, M., {et~al.} 2025, arXiv preprint:2505.09426

\bibitem[{Kapyla \& Korpi(2011)}]{KapylaZNF2010}
Kapyla, P.~J. \& Korpi, M.~J. 2011, Mon. Not. Roy. Astron. Soc., 413, 901

\bibitem[{Kiuchi {et~al.}(2015)Kiuchi, Cerd\'a-Dur\'an, Kyutoku, Sekiguchi, \& Shibata}]{Kiuchi2015KHI}
Kiuchi, K., Cerd\'a-Dur\'an, P., Kyutoku, K., Sekiguchi, Y., \& Shibata, M. 2015, Phys. Rev. D, 92, 124034

\bibitem[{Kiuchi {et~al.}(2018)Kiuchi, Kyutoku, Sekiguchi, \& Shibata}]{Kiuchi2018paper}
Kiuchi, K., Kyutoku, K., Sekiguchi, Y., \& Shibata, M. 2018, Phys. Rev. D, 97, 124039

\bibitem[{Kiuchi {et~al.}(2024)Kiuchi, Reboul-Salze, Shibata, \& Sekiguchi}]{Kiuchi_dynamo24}
Kiuchi, K., Reboul-Salze, A., Shibata, M., \& Sekiguchi, Y. 2024, Nature Astron., 8, 298

\bibitem[{Lorimer(2008)}]{Lorimer2008review}
Lorimer, D.~R. 2008, Living Rev. Rel., 11, 8

\bibitem[{Mamatsashvili {et~al.}(2013)Mamatsashvili, Chagelishvili, Bodo, \& Rossi}]{Mamatsashvili2013}
Mamatsashvili, G.~R., Chagelishvili, G.~D., Bodo, G., \& Rossi, P. 2013, Mon. Not. Roy. Astron. Soc., 435, 2552

\bibitem[{Mattia {et~al.}(2023)Mattia, Del~Zanna, Bugli, Pavan, Ciolfi, Bodo, \& Mignone}]{MattiaResRelJet}
Mattia, G., Del~Zanna, L., Bugli, M., {et~al.} 2023, Astron. Astrophys., 679, A49

\bibitem[{{Maury} {et~al.}(2022){Maury}, {Hennebelle}, \& {Girart}}]{maury22}
{Maury}, A., {Hennebelle}, P., \& {Girart}, J.~M. 2022, Frontiers in Astronomy and Space Sciences, 9, 949223

\bibitem[{McCorquodale \& Colella(2011)}]{McCorquodale:2011}
McCorquodale, P. \& Colella, P. 2011, Commun. Appl. Math. Comput. Sci., 6, 1

\bibitem[{Miravet-Ten\'es {et~al.}(2022)Miravet-Ten\'es, Cerd\'a-Dur\'an, Obergaulinger, \& Font}]{MinitMRI}
Miravet-Ten\'es, M., Cerd\'a-Dur\'an, P., Obergaulinger, M., \& Font, J.~A. 2022, Mon. Not. Roy. Astron. Soc., 517, 3505

\bibitem[{Miravet-Ten\'es \& Pessah(2025)}]{MiquelEffMRI}
Miravet-Ten\'es, M. \& Pessah, M.~E. 2025, Astron. Astrophys., 696, A2

\bibitem[{Mizuno \& Rezzolla(2025)}]{mizuno2025general}
Mizuno, Y. \& Rezzolla, L. 2025, in New Frontiers in GRMHD Simulations (Springer), 3--28

\bibitem[{Mongwane(2015)}]{Mongwane:2015}
Mongwane, B. 2015, Gen. Relativ. Gravit., 47, 1

\bibitem[{Most(2023)}]{most2023dynamoimpact}
Most, E.~R. 2023, Physical Review D, 108, 123012

\bibitem[{Palenzuela(2020)}]{palenzuela2020NR}
Palenzuela, C. 2020, Frontiers in Astronomy and Space Sciences, 7, 58

\bibitem[{Palenzuela {et~al.}(2022)Palenzuela, Aguilera-Miret, Carrasco, Ciolfi, Kalinani, Kastaun, Mi\~nano, \& Vigan\`o}]{TurbBampBNS_2022}
Palenzuela, C., Aguilera-Miret, R., Carrasco, F., {et~al.} 2022, Phys. Rev. D, 106, 023013

\bibitem[{Palenzuela {et~al.}(2018)Palenzuela, Miñano, Viganò, Arbona, Bona-Casas, Rigo, Bezares, Bona, \& Massó}]{palenzuela18}
Palenzuela, C., Miñano, B., Viganò, D., {et~al.} 2018, Class. Quantum Grav., 35, 185007

\bibitem[{Pelletier \& Pudritz(1992)}]{pelletier1992hydromagnetic}
Pelletier, G. \& Pudritz, R.~E. 1992, Astrophysical Journal, Part 1 (ISSN 0004-637X), vol. 394, no. 1, July 20, 1992, p. 117-138. Research supported by NATO and NSERC., 394, 117

\bibitem[{Price \& Rosswog(2006)}]{priceRosswog2006}
Price, D.~J. \& Rosswog, S. 2006, Science, 312, 719

\bibitem[{Radice \& Hawke(2024)}]{radice2024turbulence}
Radice, D. \& Hawke, I. 2024, Living Reviews in Computational Astrophysics, 10, 1

\bibitem[{Rayleigh(1917)}]{Rayleigh1917}
Rayleigh, L. 1917, Proc. R. Soc. Lond. A, 93, 148

\bibitem[{Rincon(2019)}]{rincon2019dynamo}
Rincon, F. 2019, Journal of Plasma Physics, 85, 205850401

\bibitem[{Rincon {et~al.}(2007)Rincon, Ogilvie, \& Proctor}]{Rincon2007_SSPdynamo}
Rincon, F., Ogilvie, G.~I., \& Proctor, M. R.~E. 2007, Phys. Rev. Lett., 98, 254502

\bibitem[{Ruiz {et~al.}(2016)Ruiz, Lang, Paschalidis, \& Shapiro}]{ruiz2016BNSjet}
Ruiz, M., Lang, R.~N., Paschalidis, V., \& Shapiro, S.~L. 2016, The Astrophysical Journal Letters, 824, L6

\bibitem[{Shakura \& Sunyaev(1973)}]{Shakura+1973}
Shakura, N.~I. \& Sunyaev, R.~A. 1973, Astron. Astrophys., 24, 337

\bibitem[{Shibata(2015)}]{shibata2015numerical}
Shibata, M. 2015, Numerical Relativity, 100 years of general relativity (World Scientific Publishing Company Pte Limited)

\bibitem[{Shu(1998)}]{shu98}
Shu, C.-W. 1998, Essentially non-oscillatory and weighted essentially non-oscillatory schemes for hyperbolic conservation laws (Springer Berlin Heidelberg), 325--432

\bibitem[{Siegel {et~al.}(2013)Siegel, Ciolfi, Harte, \& Rezzolla}]{siegel2013MRI}
Siegel, D.~M., Ciolfi, R., Harte, A.~I., \& Rezzolla, L. 2013, Physical Review D—Particles, Fields, Gravitation, and Cosmology, 87, 121302

\bibitem[{Squire \& Bhattacharjee(2014)}]{Squire2014}
Squire, J. \& Bhattacharjee, A. 2014, Astrophys. J., 797, 67

\bibitem[{Suresh \& Huynh(1997)}]{suresh97}
Suresh, A. \& Huynh, H. 1997, J. Comput. Phys., 136, 83

\bibitem[{Thorne \& Blandford(2017)}]{ThorneBlandford}
Thorne, K.~S. \& Blandford, R.~D. 2017, Modern Classical Physics (Princeton University Press)

\bibitem[{Vasil {et~al.}(2013)Vasil, Lecoanet, Brown, Wood, \& Zweibel}]{Vasil_2013}
Vasil, G.~M., Lecoanet, D., Brown, B.~P., Wood, T.~S., \& Zweibel, E.~G. 2013, Astrophys. J., 773, 169

\bibitem[{Velikhov(1959)}]{velikhov1959}
Velikhov, E. 1959, Sov. Phys. JETP, 36, 995

\bibitem[{Vigan\`o {et~al.}(2020)Vigan\`o, Aguilera-Miret, Carrasco, Mi\~nano, \& Palenzuela}]{viganoGRMHDLES}
Vigan\`o, D., Aguilera-Miret, R., Carrasco, F., Mi\~nano, B., \& Palenzuela, C. 2020, Phys. Rev. D, 101, 123019

\bibitem[{Viganò {et~al.}(2019{\natexlab{a}})Viganò, Aguilera-Miret, \& Palenzuela}]{vigano_gradientMHD}
Viganò, D., Aguilera-Miret, R., \& Palenzuela, C. 2019{\natexlab{a}}, Physics of Fluids, 31, 105102

\bibitem[{Viganò {et~al.}(2019{\natexlab{b}})Viganò, Martínez-Gómez, Pons, Palenzuela, Carrasco, Miñano, Arbona, Bona, \& Massó}]{vigano19}
Viganò, D., Martínez-Gómez, D., Pons, J., {et~al.} 2019{\natexlab{b}}, Comput. Phys. Commun., 237, 168

\end{thebibliography}

\begin{appendix}
\section{Impact of more general gradients on the MRI}\label{app:verticalgrads}

As we have seen that radial gradients in the magnetic field can have a crucial impact on the MRI, in this appendix we consider the impact of additional gradients on the instability.
Specifically, we consider the case where we included vertical gradients in the background magnetic field as well as in the velocity (on top of differential rotation).
In essence this means we took 
\begin{subequations}
\begin{align}
    &\vec v = \left[\Om(R,z)- \Om(R_0,z_0)\right](R_0+x)\hat y \;, \\
    &\vec B = B^y(z)\hat y + B^z \;, \quad \partial_iB^i = \partial_z B^z = 0 \;, 
\end{align}
\end{subequations}
so that the gradients term entering the WKB expansions are 
\begin{subequations}
\begin{align}
    & \partial_i v_j = \delta_{j2}\delta_{i1}S_0 + \delta_{j2}\delta_{i3}S_1 \;, \quad S_1 = \frac{R}{z}\left(\frac{\partial\Omega}{\partial\log z}\right)\;, \\
    & \partial_i B_j = \delta_{i3}\delta_{j2} (\partial_z B^y) \;, 
\end{align}
\end{subequations}
while $S_0$ is defined as in \cref{eq:corotatingObserver}.
We also used the no-monopoles constraint to kill gradients in the vertical magnetic field component. 

Performing the linear analysis under the same simplifying assumptions as in Sect. \ref{subsec:radialgrads}, the perturbation equations corresponding to the background configuration under consideration read 
\begin{subequations}\label{subeqs:perturbedMHDVerticalGrad}
\begin{align}
    & k_x \delta v^x + k_z \delta v^z = 0 \;, \\
     -i &\om \delta v^x - 2\Om_0\delta v^y + i k_x (v_A^y \delta v_A^y + v_A^z \delta v_A^z ) -i \delta v_A^x k_z v_A^z = 0 \;,\\
    -i &\om \delta v^y +\frac{\kappa^2 }{2\Om}\delta v^x  - i k_z v_A^z \delta v_A^y  \textcolor{blue}{+ S_1 \delta v^z - a \delta v_A^z}= 0 \;,\\
    -i &\om \delta v^z   + i k_z v_A^y \delta v_A^y \textcolor{blue}{+ a \delta v_A^y}= 0 \;, \\
    -i & \om \delta v_A^x - i v_A^z k_z \delta v^x = 0 \;,  \\
    -i &\om \delta v_A^y  - i v_A^z k_z \delta v^y - S_0 \delta v_A^x \textcolor{blue}{+ a \delta v_A^z - S_1 \delta v_A^z}= 0 \;,\\
    -i & \om \delta v_A^z  -i v_A^z k_z \delta v^z = 0\;,
\end{align}
\end{subequations}
where we have introduced $a = \partial_z v_A^y$ to slim down the notation. 
As before, we first rewrote the various perturbed quantities in terms of $\delta v^z$:
\begin{subequations}\label{subeqs:substitutionsVerticalGrad}
\begin{align}
    & \delta v^x = - \frac{k_z}{k_x}\delta v^z \;,\\
    & \delta v^y = i \left(\frac{\om}{\omega^2 - p^2}\right)\frac{k_z}{k_x}\left[\frac{\kappa^2}{2\Om} - S_0\frac{p^2}{\om^2} \textcolor{blue}{-S_1 \frac{k_x}{k_z}\left(1-\frac{p^2}{\om^2}\right)} \right]\delta v^z \;,\\
    & \delta v_A^x = \frac{k_z}{k_x} \frac{p}{\omega}\delta v^z \;,\\
    & \delta v_A^y = \frac{\om}{k_zv_A^y\textcolor{blue}{-ia}} \delta v^z\;,\\
    & \delta v_A^z = - \frac{p}{\om}\delta v^z \;,
\end{align}
\end{subequations}
and substituting these into the $x$-component of the Euler equation to arrive at
\begin{multline}
    \left(1 + \frac{k_x^2}{k_z}\frac{v_A^y}{v_A^y \textcolor{blue}{-ia}}\right)\om^4 - \left[2p^2 + p^2 \frac{k_x^2}{k_z^2} + \kappa^2 \textcolor{blue}{-2\Om S_1\frac{k_x}{kz}}+ \frac{k_x^2}{k_z}\frac{v_A^y p^2}{v_A^y k_z \textcolor{blue}{-ia}}\right] \om^2 \\ 
    + p^2 \left[2\Om S_0 \textcolor{blue}{-2\Om S_1 \frac{k_x}{k_z} } + p^2 \left(1+ \frac{k_x^2}{k_z^2}\right)\right] =0.
\end{multline}
It is easy at this point to see that if we consider vertical modes only ($k_x=0$, as in Sect. \ref{sec:extended_MRI}), the dispersion relation reduces to the standard MRI case we discussed in Sect. \ref{subsubsec:MRI}. 
This means that adding vertical gradients in the background velocity and/or magnetic field, separately or combined, does not impact on the usual MRI phenomenology and criteria. 

Without showing the details of the calculation, let us also mention that one can make progress in the case where we included radial and vertical gradients in the velocity, and also radial gradients in the magnetic field. 
In this last case in fact the dispersion relation can be reduced to the form 
\begin{multline}
    \om^4 - (2p^2 + \kappa^2 + f)\om^2 + [p^4 + 2\Om S_0 p^2 + 2i \Om S_1 p (\partial_x v_A^z) + fp^2 ] = 0 \;,
\end{multline}
while the function $f$ is defined as in \cref{eq:def_f}.
While the equation is now complex, thus preventing a simple analytical solution, it is easy to see that when considering vertical gradients in the velocity as well as radial gradients in both the velocity and the magnetic field, the solution to the approximated dispersion relation is identical to the case with radial gradients only.

All the different cases analysed here and in Sect. \ref{subsec:radialgrads} are conveniently summarized in \cref{tab:recap}, which makes it evident the central role played by radial magnetic field gradients in the generalized instability. 

\section{A closer look at the topological indicators} \label{app:SimContact}
\begin{figure*}
    \centering
    \includegraphics[width=0.9\linewidth]{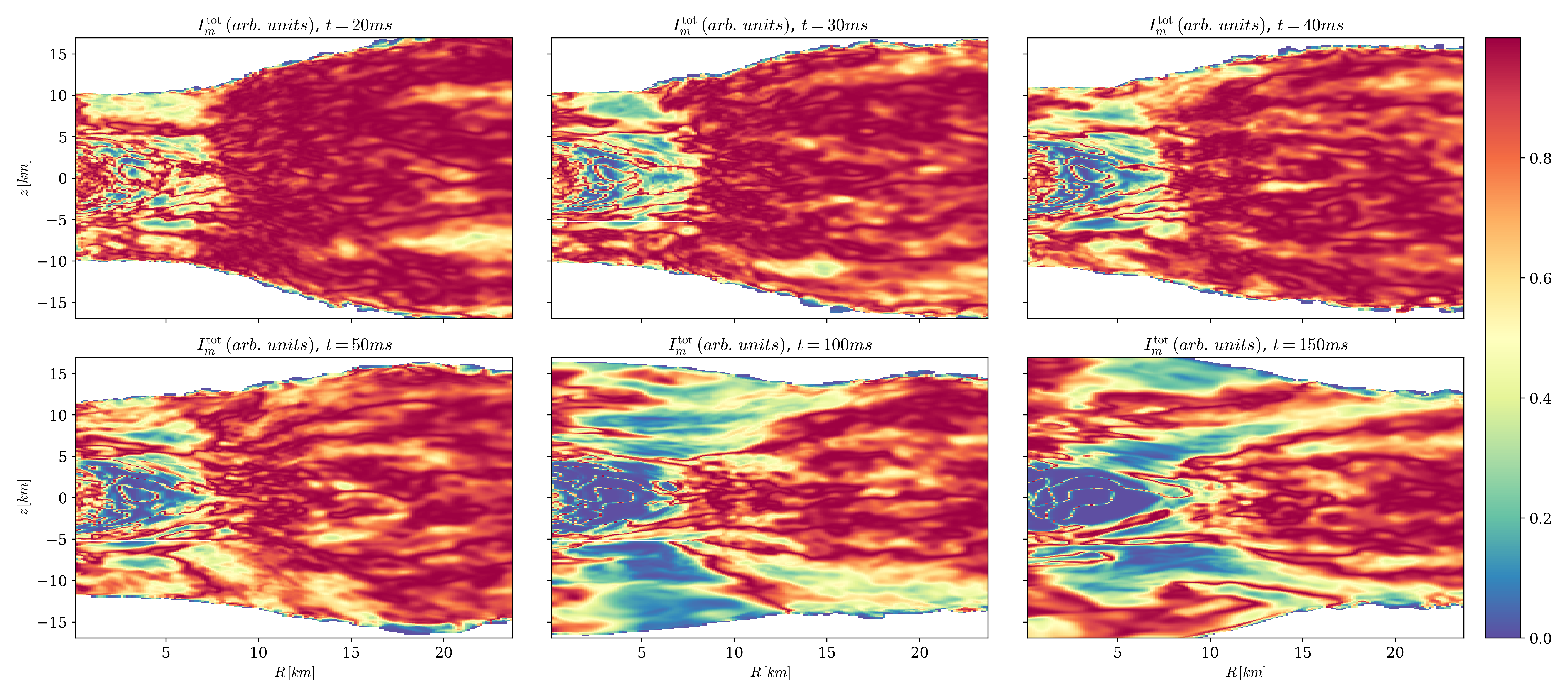}
    \caption{\textit{Evolution of the total magnetic non-axisymmetric indicators, $I_m^{\text{tot}}$}. We observe the non-axisymmetric indicator is $I_m\sim1$ even at late times $(t\gtrsim150)$ ms, particularly at large radii and close to the equatorial plane. 
    We focus here on the total indicator, which is dominated by the azimuthal contribution.
    When plotting the poloidal indicators, we observe that they are $\approx 1$ at all radii and vertical distances from the equatorial plane even at late times, $t\simeq 150$ ms.
    }
    \label{fig:NonAxisymIndEvol}
\end{figure*}
The aim of this appendix is to discuss more in detail the indications we have from the simulations presented in \cite{RoleTurbRicard2023,DelayedJet24}. 
We focus in particular on the two topological indicators that are introduced there, and can offer useful insights into the amount of turbulence in the post-merger environment. 
By means of a Fourier-type decomposition, we argue that the behaviours observed when these indicators are evaluated on the simulations output is supportive of the generalized MRI conditions we studied in this work. 

The non-axisymmetry indicators are introduced as follows:
\begin{equation}
    I_K^i = \frac{ \delta E^i}{\bar E^{i}_K  + \delta E^i}\;, \quad I_m = \frac{ \delta E^i_m}{\bar E^i_m + \delta E^i_m}\;,
\end{equation}
where 
\begin{subequations}
\begin{align}
     \bar E^i_K &= \bar U^i_K \bar U_i^K \;, \quad \delta  E^i_K = \delta U^i_K \delta U_i^K \;, \\
     \bar E^i_m &= \bar U^i_m \bar U_i^m \;, \quad \delta  E^i_m = \delta U^i_m \delta U_i^m \;,
\end{align}
\end{subequations}
and the two vector fields considered are $U^i_K = \sqrt{\rho}v^i$ and $U^i_m = B^i$. 
These vector fields (components) are decomposed into an average along the azimuthal direction (cf. \cref{eq:azimuthal_average} in main text) and an integrated residual, 
\begin{align}
    \delta U^i ( R,z) &= \frac{\int_l|U^i - \bar U^i |\chi^{-1/2}Rd\varphi}{\int_l \chi^{-1/2}Rd\varphi} \;.
\end{align}
Evaluating these indicators on the simulations output---considering either the dominant azimuthal component only or the summed contributions of the individual components---the authors find the kinetic axisymmetric indicator to be $I_K \simeq 0.05$ at most---thus showing a high degree of axisymmetry---while the magnetic one $I_m \sim 1$. 
This is true both at early ($t\gtrsim 20 $ms) and late times ($t\gtrsim150$ms) and at all radii $R \gtrsim 10$km.
At the same time though, the residual energies are similar in values $\delta E_K^\varphi\simeq \delta E_m^\varphi$ at all radii, and until late times $t\sim 150$ ms. 

To unpack the information encoded in the indicators, we introduced a Fourier decomposition in the azimuthal direction, 
\begin{equation}
    U^\varphi = c_0(R,Z) + \sum_{n=1}^\infty a_n(R,z) \cos(n\varphi)  + \sum_{n=1}^\infty b_n(R,z) \sin(n\varphi)  \;, 
\end{equation}
and rewrote the averaged energy as $\bar E^\varphi = (c_0)^2$.
As for the residual, we can write the following inequality,
\begin{equation}
    \delta U^\varphi \leq  \sum_{n=1}^{\infty} \frac{2}{\pi}(|a_n| + |b_n|) \;,
\end{equation}
so that for the purpose of the following argument we can write 
\begin{equation}
    \delta E^\varphi \simeq \left(\sum |a_n| + |b_n|\right)^2 \;, \qquad I \simeq \left[1 + \frac{(c_0)^2}{\left(\sum |a_n| + |b_n|\right)^2}\right]^{-1}\;.
\end{equation}
In terms of this decomposition we can make the following observations
\begin{enumerate}
    \item As $I_K\ll 1$ the fluid motion is mostly axisymmetric, and for the purpose of a linear analysis the background velocity at $t\gtrsim 20$ms can be obtained using azimuthal averages. 
    \item Since $I_m\simeq 1$ but $\delta E_K \simeq \delta E_m$, we see that the difference in the axisymmetric indicators really comes from the fact that the averaged kinetic energy is dominating over the magnetic one---the background is in a `kinematic regime' in a loose sense.
    \item Since $I_m\simeq 1, \delta E_K \simeq \delta E_m$ at all radii and up until late times (cf.  Fig. $5$b of \citealt{DelayedJet24}), the azimuthal average of the magnetic field varies (in the radial direction) on the same scales as the residuals in the velocity. 
\end{enumerate}
In short then, the indicators suggest that in order to analyse the role of the MRI in the simulations, we first need to generalize the standard criteria to the case where we considered (radial) gradients in the magnetic field on top of differential rotation, as discussed in Sect. \ref{sec:extended_MRI}.

In this work we often considered 2D maps of the relevant averaged quantities, showing them as a function of the cylindrical coordinates $(R,z)$. 
As such, we here show in \cref{fig:NonAxisymIndEvol} the evolution of the total magnetic non-axisymmetric indicator at some representative times---which is dominated by the azimuthal component. 
The figure clearly shows that the non-axisymmetric indicator is close to $1$ in large parts of the domain even at late times $\mathcal{O}(100)$ ms. 
Effectively, this raises an additional caveat on the applicability of the MRI criteria in the simulation considered here. 
We also note a similar point may apply to some protoplanetary disks where no ordered large-scale field is observed \citep{huang24,huang25}.

\section{Empirical quality factors} \label{app:QualityFactors}

\begin{figure*}
    \centering
    \includegraphics[width=0.9\linewidth]{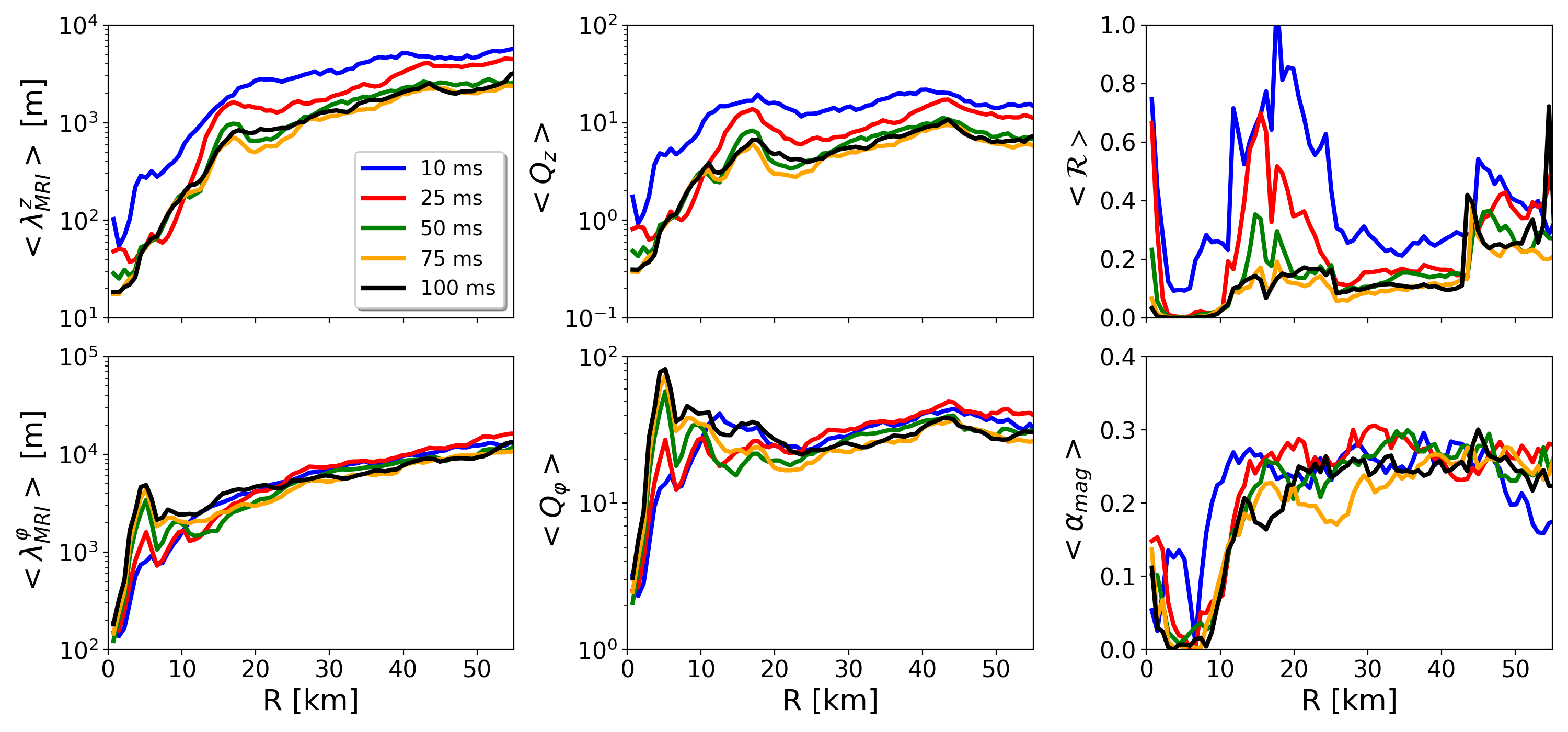}
    \caption{\textit{Quantities of empirical criteria.} \textit{Top}: Fastest-growing MRI vertical wavelength ($\lambda^z_{MRI}$), its corresponding quality factor ($Q_z$), and the ratio 
    ${\cal R} = B_{R}^2 / B_{\varphi}^2$. \textit{Bottom}: Fastest-growing MRI toroidal wavelength ($\lambda^\varphi_{MRI}$), toroidal quality factor ($Q_\varphi$), and the ratio of Maxwell stress
    to magnetic pressure ($\alpha_{mag}$). All these quantities are shown as a function of distance (in km) for different times after the merger.}
    \label{fig:alphamag}
\end{figure*}

In this appendix we present several quantitative diagnostics that have been established as useful empirical criteria for assessing the adequacy of a simulation in capturing MRI-driven MHD turbulence \citep{HawleyConvergence2011,HawleyConvergence2013}.
Specifically, we focus on the vertical and azimuthal quality factors $Q_z$ and $Q_{\varphi}$, defined as the ratio of the corresponding characteristic MRI wavelength ($\lambda_{\rm MRI}$) to the grid spacing in the vertical and azimuthal directions, respectively. Following \cite{Kiuchi2018paper}, we also examined the ratio of Maxwell stress to magnetic pressure $\alpha_{\rm mag}$, as well as the ratio between the radial and toroidal magnetic components ${\cal R} = B_{R}^2 / B_{\varphi}^2$. These quantities are plotted as functions of radius at some representative snapshots after the merger.

As shown in \cref{fig:alphamag}, the vertical quality factor remains around $Q_z \sim 8$–$10$ at radii larger than $10$–$15$km, while the azimuthal quality factor reaches $Q_\varphi \gtrsim 30$. Although the vertical quality factor suggests the resolution is only marginally sufficient, the higher values of $Q_\varphi$ are expected to compensate for the modest $Q_z$, consistent with previous convergence studies.
We also show the radial profile of $\alpha_{\rm mag}$, finding that it saturates at approximately $0.25$-$0.3$ shortly after the merger and remains roughly constant thereafter. 
Overall, the computed global quantities are consistent with —though slightly lower than—  those reported in the literature when simulations are performed at sufficiently high resolution (see e.g. \citealt{Kiuchi2018paper}). 
In particular, we note that the two main differences between our simulations and those in the literature are: (i) we started with a significantly weaker (and more realistic) initial magnetic field, and (ii) we employed LES techniques with the gradient sub-grid-scale model to capture more accurately the magnetic field amplification during the turbulent phase.

We also stress that in this appendix the quality factors $Q_z$ and $Q_\varphi$ as a function of the radial distance $R$ have been computed  using the standard MRI wavelength $\lambda_{\rm MRI}$ to enable a clearer comparison with results reported in the literature. It is reasonable to expect that generalized quality criteria based on the extended MRI wavelengths will also provide useful information on the adequacy of a simulation, since $\lambda_{\rm eMRI}$ reduces to $\lambda_{\rm MRI}$ in the absence of significant gradients and these criteria are derived from linear analysis.
The situation is different for the remaining diagnostics considered, $\alpha_{\rm mag}$ and $\mathcal{R}$, which are directly linked to the non-linear development of MHD turbulence \citep{HawleyConvergence2011}. 
As the impact on the characteristic wavelength is moderate compared to that on the timescale, we expect the quality factors to be moderately smaller, which may help explain the slightly lower values observed for the computed global quantities.
Finally, we note that smaller-scale magnetic fields, such as those generated during a Kelvin--Helmholtz amplification phase, are more susceptible to numerical dissipation, which is also known to impact on the efficiency of MRI-driven dynamos in local simulations \citep[e.g.][]{FromangZNF2007}.

\end{appendix}
\end{document}